\documentstyle[12pt,epsfig]{article}

\begin{document}

\renewcommand{\theequation}{\thesection .\arabic{equation}}
\renewcommand{\section}[1]{\setcounter{equation}{0}
 \addtocounter{section}{1}
 \vspace{5mm} \par \noindent {\large \bf \thesection . #1}
 \setcounter{subsection}{0} \par \vspace{2mm} } 
\newcommand{\sectionsub}[1]{\addtocounter{section}{1}
 \vspace{5mm} \par \noindent {\bf \thesection . #1}
 \setcounter{subsection}{0}\par}
\renewcommand{\subsection}[1]{\addtocounter{subsection}{1}
 \vspace{2.5mm}\par\noindent {\bf  \thesubsection . #1}\par
 \vspace{0.5mm}}
\renewcommand{\thebibliography}[1]{{\vspace{5mm}\par 
 \noindent{\large \bf References}\par \vspace{2mm}} 
 \list{$[$\arabic{enumi}$]$}{\settowidth\labelwidth{[#1]}
 \leftmargin \labelwidth \advance\leftmargin\labelsep
 \addtolength{\topsep}{-4em}\usecounter{enumi}}
 \def\newblock{\hskip .11em plus .33em minus .07em}
 \sloppy\clubpenalty4000\widowpenalty4000 \sfcode`\.=1000\relax 
 \setlength{\itemsep}{-0.4em}}
\newcommand{\acknowledgments}[1]{\vspace{5mm}\par 
 \noindent{\large \bf Acknowledgments}\par \vspace{2mm}}

\def\a{& \hspace{-7pt}}
\def\bea{\begin{eqnarray}}
\def\eea{\end{eqnarray}}
\def\be{\begin{equation}}
\def\ee{\end{equation}}
\def\nn{\nonumber}
\def\alp{\alpha}
\def\bet{\beta}
\def\gam{\gamma}
\def\del{\delta}
\def\eps{\epsilon}
\def\sig{\sigma}
\def\lam{\lambda}
\def\Lam{\Lambda}
\def\m{\mu}
\def\n{\nu}
\def\r{\rho}
\def\s{\sigma}
\def\d{\delta}
\def\Z{{\bf Z}}
\def\e{\epsilon}
\def\st{\scriptstyle}
\def\mco{\multicolumn}
\def\epp{\epsilon^{\prime}}
\def\vep{\varepsilon}
\def\ra{\rightarrow}
\def\ab{\bar{\alpha}}

\newcommand{\sect}[1]{\setcounter{equation}{0} \section{#1}}
\newcommand{\eqn}[1]{(\ref{#1})}
\newcommand\rf[1]{(\ref{#1})}
\newcommand{\NPB}[3]{{Nucl.\ Phys.} {\bf B#1} (#2) #3}
\newcommand{\CMP}[3]{{Commun.\ Math.\ Phys.} {\bf #1} (#2) #3}
\newcommand{\PRD}[3]{{Phys.\ Rev.} {\bf D#1} (#2) #3}
\newcommand{\PLB}[3]{{Phys.\ Lett.} {\bf B#1} (#2) #3}
\newcommand{\JHEP}[3]{{JHEP} {\bf #1} (#2) #3}
\newcommand{\ft}[2]{{\textstyle\frac{#1}{#2}}\,}
\newcommand{\dt}{\partial_{\langle T\rangle}}
\newcommand{\dtbar}{\partial_{\langle\bar{T}\rangle}}
\newcommand{\al}{\alpha^{\prime}}
\newcommand{\mst}{M_{\scriptscriptstyle \!S}}
\newcommand{\mpl}{M_{\scriptscriptstyle \!P}}
\newcommand{\dv}{\int{\rm d}^4x\sqrt{g}}
\newcommand{\lv}{\left\langle}
\newcommand{\rv}{\right\rangle}
\newcommand{\ph}{\varphi}
\newcommand{\sbar}{\,\bar{\! S}}
\newcommand{\xbar}{\,\bar{\! X}}
\newcommand{\fbar}{\,\bar{\! F}}
\newcommand{\zbar}{\,\bar{\! Z}}
\newcommand{\tbar}{\bar{T}}
\newcommand{\ybar}{\bar{Y}}
\newcommand{\phb}{\bar{\varphi}}
\newcommand{\cm}{Commun.\ Math.\ Phys.~}
\newcommand{\pr}{Phys.\ Rev.\ D~}
\newcommand{\prl}{Phys.\ Rev.\ Lett.~}
\newcommand{\pl}{Phys.\ Lett.\ B~}
\newcommand{\ibar}{\bar{\imath}}
\newcommand{\jbar}{\bar{\jmath}}
\newcommand{\np}{Nucl.\ Phys.\ B~}
\newcommand{\eqalign}[1]{\begin{array}{ll} #1 \end{array}}
\newcommand{\gsi}{\,\raisebox{-0.13cm}{$\stackrel{\textstyle>}
 {\textstyle\sim}$}\,}
\newcommand{\lsi}{\,\raisebox{-0.13cm}{$\stackrel{\textstyle<}
 {\textstyle\sim}$}\,}
\newcommand{\ds}[1]{\displaystyle{#1}}

\thispagestyle{empty}

\begin{center}
\hfill UvA-WINS-Wisk-99-10 \\
\hfill LMU-TPW 99-13\\
\hfill{\tt hep-th/9907112}\\

\vspace{1.7cm}

{\Large\bf Anomaly cancellation in K3 orientifolds}\\

\vspace{1.4cm}

{\sc Claudio A. Scrucca$^{a,1}$ and Marco Serone$^{~b,c,2}$} \\

\vspace{1.2cm}

${}^a$
{\em Department of Physics, Ludwig Maximilian University of Munich}\\
{\em Theresienstra\ss e 37, 80333 Munich, Germany}\\

\vspace{.3cm}

${}^b$
{\em Department of Mathematics, University of Amsterdam}\\
{\em Plantage Muidergracht 24, 1018 TV Amsterdam, The Netherlands} \\

\vspace{.3cm}

${}^c$
{\em Spinoza Institute, University of Utrecht} \\
{\em Leuvenlaan 4, 3584 CE Utrecht, The Netherlands} \\

\end{center}

\vspace{0.8cm}

\centerline{\bf Abstract}
\vspace{2 mm}
\begin{quote}\small

We study in detail the pattern of anomaly cancellation in $D=6$ Type IIB $\Z_N$ 
orientifolds, occurring through a generalized Green-Schwarz mechanism involving 
several RR antisymmetric tensors and scalars fields. 
The starting point is a direct string theory computation of the inflow of anomaly 
arising from magnetic interaction of D-branes, O-planes and fixed-points, 
which are encoded in topological one-loop partition functions in the RR odd
spin-structure. All the RR anomalous couplings of these objects are then 
obtained by factorization. They are responsible for a spontaneous
breaking of $U(1)$ factors through a Higgs mechanism involving the corresponding
hypermultiplets. Some of them are also related by supersymmetry to gauge 
couplings involving the NSNS scalars sitting in the tensor multiplets. We also 
comment on the possible occurrence of tensionless strings when these couplings 
diverge.

\end{quote}

\vfill
\begin{flushleft}
\rule{16.1cm}{0.2mm}\\[-3mm]
\end{flushleft}
$^1${\footnotesize \tt Claudio.Scrucca@physik.uni-muenchen.de}\\
$^2${\footnotesize\tt serone@wins.uva.nl}

\newpage
\setcounter{equation}{0}


\section{Introduction}

Anomalies have proven to play a prominent role in the study of non-trivial
vacua of string theory. Their cancellation constitute a very severe constraint 
on the low-energy quantum field theory, which can be usually understood in string 
theory as consequence of basic consistency requirements, like modular invariance 
and tadpole cancellation. A very interesting 
peculiarity of low-energy effective quantum field theories emerging from string 
theory is the occurrence of anomalies in certain tree-level magnetic interaction 
mediated by p-form gauge fields, beside the well known anomalies arising in one-loop 
amplitudes involving chiral fermions or self-dual bosons. This allows for consistent 
quantum fields theories with a non-vanishing one-loop anomaly cancelled by an equal 
and opposite tree-level anomaly. This is of course the celebrated Green-Schwarz
mechanism \cite{GS}.

Thanks to the important recent developments in string theory, this cancellation 
mechanism can be understood as a particular example of the so-called inflow 
mechanism \cite{ch}. Given a consistent anomaly free theory, there can exist vacua
with topological defects supporting anomalous zero modes. The latter will be 
responsible for a local violation of charge conservation on the defects, which
has to be compensated by an inflow of charge from outside the defects induced
by appropriate couplings of the defects to the fields living in the bulk.
In string theory, this kind of non-perturbative objects do indeed arise in 
non-trivial vacua, important examples being D-branes and O-planes
\cite{pol}. Both of these kind of objects support anomalous fields on their 
world-volumes, and have thus to have appropriate anomalous couplings to bulk fields
for consistency. Indeed, they present topological Wess-Zumino couplings to RR p-forms
which are precisely those required to cancel through the inflow mechanism all
the one-loop anomalies on their world-volumes \cite{ghm,cy,ss}. The effective appearance
of all of these couplings in string theory has been deduced in a unified way in 
\cite{mss} by factorizing RR magnetic interactions between D-branes and O-planes
with non trivial curvatures, encoded in annulus, M\"obius strip and Klein bottle 
amplitudes. They have also been checked in successive steps through direct
computations of the induced RR charges, encoded in disk and crosscap amplitudes
\cite{li,dou,cr,stef}. Some terms have also been derived through duality 
\cite{bsv,djm,dm}. 
In orientifold models \cite{sag,ps,hor,dlp}, the situation is particularly clear.
The D-branes associated to Chan-Paton degrees of freedom,
as well as the fixed-planes of the orientifold group action like O-planes, support in 
general anomalous fields giving rise to a non-vanishing anomaly, but also appropriate 
Wess-Zumino couplings to RR fields inducing a net inflow of anomaly. 
For consistent models, these one-loop and tree-level anomalies always cancel; this is 
usually a direct consequence of the more basic consistency requirement of tadpole 
cancellation.

In this paper, we study in detail anomaly cancellation in Type IIB $\Z_N$ orientifolds 
in $D=6$ dimensions \cite{ps,bs,GP,GJ,DP}. These models are particularly interesting because
of their rich and peculiar spectrum, which differs from what expected \cite{pol2} for a 
smooth Type IIB orientifold on $K3$, which is equivalent to Type I on $K3$ as a 
consequence of the well known fact that Type I strings can be understood as the simplest 
orientifold of Type IIB strings \cite{sag}.
In particular, a variable number of extra tensor multiplets arise in addition to the
universal one expected in the smooth case \cite{bs}. Anomaly cancellation imposes extremely 
powerful constraints on these theories, like on every $N=(1,0)$ supersymmetric chiral 
theory in $D=6$, due to the potential presence of both gauge and gravitational anomalies 
\cite{agw}. In general, a generalization of the Green-Schwarz mechanism involving all
the RR fields present in the model occurs \cite{sagGS}. In \cite{ss}, a systematic way of 
studying anomaly cancellation in orientifold models was sketched, and applied to the simple 
case of the $\Z_2$ model, investigated in great detail in \cite{americani}, in which
no extra tensor multiplets arise\footnote{Anomaly cancellation in $D=4$ $N=1$ orientifolds 
has been studied in \cite{iru}.}. 
This work is devoted to a detailed analysis of all
the $\Z_N$ models, with particular emphasis on the inflow interpretation, for the case of 
maximal unbroken gauge group. We will proceed
with a refinement of the philosophy used in \cite{mss,ss}, consisting in a direct 
string computation of the inflow of anomaly arising from magnetic interaction among 
D-branes, O-planes and fixed-points. All the anomalous couplings in each model can then
be easily obtained by factorization. Due to supersymmetry, these give an important
information on other terms of the low-energy effective action, in particular to the form 
of the gauge couplings $1/g^2$ present in the theory \cite{sagGS}.
Interestingly, there are special points in the moduli 
space of the tensor multiplet scalars where $1/g^2$ vanishes or becomes negative 
\cite{sagGS}; tensionless strings appear then in the spectrum and the theory may undergo
a phase transition \cite{dmw,sw,dlpo}. 
The orientifold models discussed here correspond to particular points in the moduli space 
parametrized by these scalars, where all the gauge couplings are positive definite and 
finite. However, from some coefficients appearing in the anomaly polynomial and by 
supersymmetry, one can find the dependence of the gauge couplings on these moduli. 
We find points where these are vanishing and no simple string description is available.
One can also fix the couplings between these scalars and the gauge fields. 
The results are in agreement with what found in \cite{abd}, where such couplings were 
computed for the D9-brane gauge group by factorizing one-loop CP-even 
amplitudes\footnote{See again \cite{iru} for similar couplings in $D=4$ $N=1$ models.}.

Analogously to what has been shown for the ${\bf Z}_2$ orientifold in \cite{americani},
anomaly cancellation involves also the exchange of scalar fields and their dual 
four-forms. This results in the breaking of some (but in general not all, even in the 
case of maximal gauge symmetry discussed here) of the $U(1)$ gauge factors 
present in the gauge group. 

This paper is organized as follows. In section two, we explain the strategy used to 
compute the inflow of anomaly through one-loop computations in the RR odd spin-structure.
In section three, we apply the general results obtained in section two to the $\Z_N$
orientifolds. In section four, we find the anomalous couplings of D-branes, O-planes and
fixed-points by factorization, and in section five we study explicitly the factorization
of the anomaly for each model. In section six, we analyze the form of some terms in the 
low-energy effective action of these models, related by supersymmetry to the anomalous 
couplings found previously, and the spontaneous breaking of some $U(1)$ factors 
of the gauge group. Finally, in last section, we give some conclusions and in the 
two appendices we respectively recall useful formulae about the inflow mechanism 
and report the explicit form of the anomalous couplings for each model.

\section{Inflow and anomalies in string theory}

One-loop anomalies in quantum field theory as well as in string theory 
can be computed by evaluating one-loop diagrams with suitable 
external particles (gravitons and gauge fields for the case of gravitational 
and gauge anomalies), one of which polarized longitudinally. 
Consistent string theories are known to be anomaly-free. 
At the level of their low-energy effective actions, the total one-loop anomaly 
need not to vanish, but can cancel against anomalies due to particular couplings 
between gauge particles and RR fields.
The crucial feature of these actions is that these couplings do actually appear
and precisely in the appropriate form to give an anomaly free quantum field theory. 
One could be satisfied by this statement and conclude that such couplings 
can be derived by requiring the tree-level
anomaly they produce to be equal and opposite to the one-loop anomaly produced 
by the chiral fields present in the model. Such point of view is however 
unsatisfactory for at least two reasons. First, it is indirect and always relies 
on the assumption (generically taken to be valid) that consistent string theories 
are anomaly free. Most importantly, it does not 
allow for a microscopic analysis of the origin of such couplings in string theory.
This second reason is crucial for our purposes of computing anomalous couplings
in IIB orientifolds. It is expected, for instance, that in such models several tensor 
and hyper multiplets, coming from different twisted closed string sectors, play a 
decisive role in the inflow mechanism.
Whereas a string derivation of these couplings allows a total understanding
of their origin and the role they play in the various inflows, an analysis at the 
level of the low-energy effective action only, would be necessarily incomplete.

The question is then how to compute such anomalous couplings (or the corresponding 
inflows) in string theory, and more precisely in generic Type IIB orientifolds. 
The most direct approach would be to perform a disk or crosscap
computation with the insertion of gauge fields, gravitons and the RR field
in question. The advantage of this approach is that it yields directly the 
couplings. However, the precise form of the vertex operators of the various tensor 
fields is needed, including those arising from the closed string twisted sectors,
and it is quite awkward and hard to fix correctly the normalizations. 
On the other hand, one can compute appropriate amplitudes from which such couplings 
can be extracted by factorization. The most convenient choice turns out to be a 
generic one-loop correlation function in the odd spin-structure with external 
gravitons and gauge fields, one of which polarized longitudinally. These amplitudes
on the annulus, M\"obius strip and Klein bottle can indeed be factorized as a 
tree-level exchange of closed strings between two disk or crosscap sources.
As we shall argue in the following, this amounts to compute directly the anomalous 
inflows induced by these couplings. 

Before entering into the details of the computation, we would like to do some 
remarks and anticipate some results. The correlation functions above will turn out 
to be a total derivative in the moduli space of the various relevant surfaces:
annulus, M\"obius strip and Klein bottle. As well known, these surfaces have two
equivalent interpretations, either as one-loop amplitudes of open (for the annulus
and M\"obius strip) or closed (for the Klein bottle) strings, or as tree-level 
closed string amplitudes.
Correspondingly, the modular parameter of these surfaces can be taken either as
the proper time $t$ in the loop channel or the proper time $l\sim 1/t$ in the tree
channel. The inflow due to massless RR fields will arise from the IR region 
$l\rightarrow \infty$ of the tree-channel, corresponding to the UV region 
$t\rightarrow 0$ of the loop channel, where indeed usual anomalies can arise.

Due to the topological nature of the amplitudes in question, the computation
always reduces to the simple evaluation of the partition function of a 
supersymmetric quantum mechanical model. 
{}From now on, we will concentrate for simplicity on the case of six non-compact 
dimensions, since this is the case we are interested in. More general cases can 
be treated similarly without additional difficulties, the fields along the compact 
directions entering always through their odd spin-structure partition
function. We are then interested to compute a given four-point function of gluons 
and/or gravitons in the odd spin-structure, with one of them polarized 
longitudinally. The unphysical particle represents essentially a gauge 
transformation, and choosing it to be a gluon or a graviton corresponds to compute 
gauge and gravitational anomalous variations respectively. 

In the odd spin-structure, 
the genus one world-sheets we are considering admit a gravitino zero mode. 
Independently of the boundary conditions, the integration over the latter in the 
Polyakov path-integral results in the insertion of the sum of the left and right 
moving world-sheet supercurrents, $T_F + \tilde{T}_F$.
Moreover, since the total superghost charge is 1, one has to take one vertex operator
in the ($-1$)-picture and all the others in the (0)-picture. A
non-vanishing result is obtained by taking the vertex of the unphysical
particle to be in the ($-1$)-picture and all the vertices of the physical 
particles in the (0)-picture. We have then to compute
\be
C= \langle\,V^{phy.}_1\,V^{phy.}_2\,V^{phy.}_3\,
V^{unphy.}\,(T_F + \tilde{T}_F)\,\rangle \;.
\label{A}
\ee
In the following, we will use the same strategy as in the introduction of 
\cite{polcai}. For simplicity, we will systematically omit the ghost and 
superghost dependence of all the operators, since their contributions to the 
correlation functions we will compute always cancel by supersymmetry.

The vertex operators in the (0)-picture for physical gluons and gravitons, with 
transverse polarizations $\epsilon^a_{\mu}$ and $\xi_{\mu\nu}$, are the usual ones. They
are given by
\bea
V^{phy.}_\gamma \a=\a \xi_\mu^a\,\lambda^a\,\oint \! d\tau\, 
\left(\dot X^\mu + i p \cdot \psi \psi^\mu \right)\,e^{ip\cdot X} \;,\\
V^{phy.}_g \a=\a \xi_{\mu\nu}\, \int\! d^2z\, 
\left(\partial X^\mu + i p \cdot \psi \psi^\mu \right)
\left(\bar \partial X^\nu + i p \cdot \tilde \psi \tilde \psi^\nu \right)\,
e^{ip\cdot X} \;.
\eea
The vertex operators in the ($-1$)-picture for unphysical gluons and gravitons, with 
longitudinal polarizations $\epsilon^a_{\mu} = p_{\mu}\,\eta^a$ and 
$\xi_{\mu\nu} = p_{(\mu} \eta_{\nu)}$, are not total derivatives as they would 
be in the (0)-picture. However, they can be written as the world-sheet supersymmetry 
variations
\be
V^{unphy.} = [Q + \tilde Q, \hat V^{unphy.}]
\ee
of the auxiliary vertices 
\bea
\hat V^{unphy.}_\gamma \a=\a -i\,\eta^a\,\lambda^a\,\oint\!d\tau\,e^{ip\cdot X} \;,
\label{unpgl} \\
\hat V^{unphy.}_g \a=\a -2i\,\eta_\mu\,\int\! d^2z\,  
\left[(\partial + \bar \partial) X^\mu + 
i p \cdot (\psi - \tilde \psi) (\psi - \tilde \psi)^\mu \right] 
\,e^{ip\cdot X}\;. \label{unpgr}
\eea
In the above expressions, $\lambda^a$ are the Chan-Paton matrices, and 
$$
Q = \oint_{C_\tau}\! \frac {dz}{2 \pi i}\, j \;,\;\;
\tilde Q = \oint_{C_\tau}\! \frac {d \bar z}{2 \pi i}\, \tilde j \;,
$$
are the left and right-moving world-sheet supercharges.

We can now use standard manipulations as in \cite{fms} and reverse the contour 
integration for $Q$ and $\tilde Q$\footnote{In the case at hand, these manipulations 
have to be taken on the covering torus of each surface.}. 
Since all the other vertex operators in the correlation are supersymmetric, the only 
non-vanishing term arises when $Q + \tilde Q$ hits the picture changing vertex 
operator $T_F + \tilde T_F$, giving
\be
[Q + \tilde Q,T_F+\tilde{T}_F] = T_B+\tilde{T}_B \;,
\label{Tb}
\ee
where $T_B+\tilde{T}_B$ is the world-sheet energy-momentum tensor. 
We are then left with
\be
C = \langle\,V^{phy.}_1\,V^{phy.}_2\,V^{phy.}_3\,
\hat V^{unphy.}\,(T_B+\tilde{T}_B)\,\rangle \;.
\label{Aeff}
\ee

The insertion of the energy-momentum tensor in a correlation function corresponds 
generically to take a Teichm\"uller deformation of the remaining amplitude; in our 
case, this means basically to take a derivative with respect to the modulus
$t$. By a careful treatment of the path-integral measure, it has been 
shown that this is indeed precisely what happens for the six-gluon correlation 
function giving the hexagon gauge anomaly in type I string theory \cite{ikk}. 
It is not difficult to convince oneself that the same analysis can be carried out
in the present case without any major change, the compact space entering rather 
trivially in (\ref{Aeff}). The final result for the integrated amplitude is then
\be
A = \int_0^{\infty}\! dt \frac{d}{dt} \langle\,V^{phy.}_1\,V^{phy.}_2\,V^{phy.}_3\,
\hat V^{unphy.}\,\rangle \;.
\label{Aefff}
\ee
This is indeed in agreement with the expectation that space-time anomalies in 
string theory arise as boundary terms in the moduli space of the various
involved surfaces \cite{fms}. This limit involves also the position
of the graviton vertices. A potential anomaly can arise only when all gravitons
approach the same world-sheet boundary where all the gauge fields sit or,
for pure gravitational anomalies, when the four gravitons are on the same boundary or
crosscap \cite{liu}. These are the only kinematical configurations in which, along 
the whole one-loop surface, there is no flow of momentum $q$ in the tree-level channel
leading to an exponential suppression of the amplitude like $e^{-q^2/t}$. This is nothing
but a generalization of the argument used to show that non-planar diagrams,
in the hexagon gauge anomaly of type I, are anomaly-free (see e.g. \cite{GSW2}).
The boundary value we are interested in is 
$t\rightarrow 0$, that is $l \rightarrow \infty$, where the amplitude is dominated 
by the exchange of massless closed string states. The limit $t \rightarrow \infty$ 
corresponds instead to the IR part of the loop channel.

The full evaluation of (\ref{Aefff}), in particular when more gravitons are present,
requires still a considerable amount of work. However, it is not our aim
to perform such a calculation, that in a consistent string model has to give
necessarily a vanishing result\footnote{See for example \cite{liu}, 
where the mixed and gravitational
anomalies are computed in type I string theory, using the covariant operatorial
formalism, and are shown to vanish.}. Rather, we are interested to extract from 
(\ref{Aefff}) the various anomalous inflows coming from the exchange of RR
fields. In order to do that, we can restrict the evaluation of (\ref{Aefff}) to the
lowest order in the external momenta $p_\mu$. Indeed, in quantum field
theory, any momentum dependence in an anomalous variation of the effective action
can be removed by adding suitable local counter terms\footnote{Corrections involving
higher powers of $p_\mu$ would be in any case sub-leading in the low-energy limit
$\al \rightarrow 0$.}.
This is not possible in string theory, where the full anomalous variation
presents indeed a dependence on the external momenta, dependence that turns out to 
be crucial to get a total vanishing result. By neglecting this dependence
we will then be able to extract a finite and non-vanishing result corresponding to 
the inflow terms. Notice that in this approximation the amplitude (\ref{Aefff}) is 
$t$-independent.

Recall now that the odd spin-structure one-loop correlation function above 
involves an integration over the six bosonic zero modes $x^\mu_0$, corresponding
to the integral over six-dimensional space-time, as well as an integration over
the six fermionic zero modes $\psi_0^\mu$, which is vanishing unless all of them 
are inserted.
Notice that none of the unphysical vertices (\ref{unpgl}) and (\ref{unpgr}) 
contains fermionic zero modes, so that all of them must be provided by the physical 
vertices $V^{phy.}$. As discussed in \cite{mss}, the gluon and graviton vertex 
operators can soak at most two fermionic zero modes, and their effective form at 
leading order in the momentum $p^\mu$ is
\bea
V^{eff.}_\gamma (F) \a=\a \oint\!d\tau\,F^a\,\lambda^a\;,\\
V^{eff.}_g (R) \a=\a \int\! d^2z\,R_{\mu\nu} 
\left[X^\mu(\partial + \bar \partial) X^\mu + 
(\psi - \tilde \psi)^\mu (\psi - \tilde \psi)^\nu \right]\;,
\eea
in terms of the gauge and gravitational curvature two-forms
\bea
F^a = \frac 12\, F_{\mu\nu}^a \, \psi_0^\mu \psi_0^\nu \;,\;\;
R_{\mu\nu} = \frac 12\, R_{\mu\nu\rho\sigma}\, \psi_0^\rho \psi_0^\sigma\;.
\eea
Interestingly enough, the unphysical gluon and graviton vertex operators 
reduce, at leading order in the momentum $p^\mu$, to effective vertices 
of exactly the same kind, but with unphysical longitudinal curvatures
$F^a = \eta^a$ and $R_{\mu\nu} = p_{[\mu}\eta_{\nu]}$ which do not depend 
on the fermionic zero modes. 

Since all the effective vertices are now at most quadratic in the fields,
it is convenient to exponentiate them adding the corresponding interaction
to the free world-sheet action. The net effect of the unphysical vertex 
is to shift the corresponding curvature as $F^a \rightarrow F^a + \eta^a$
if it is a gluon, or $R_{\mu\nu} \rightarrow R_{\mu\nu} + 
{\cal D}_{[\mu}\eta_{\nu]}$ if it is a graviton. 
The integral over the six fermionic zero modes
automatically selects the correct number (three) of physical vertices or,
in language of forms, the six-form polynomial to be integrated over the
six-dimensional space-time we are considering. However, in order to get 
a single unphysical vertex, one has to take the functional derivative 
with respect to the parameter $\eta^a$ or $\eta^\mu$, and then set the 
parameter to zero. A striking similitude then emerges with Fujikawa's 
method of computing anomalies, as in {\cite{agw}}. 
The unphysical vertex plays the role of the anomalous 
variation of the path-integral measure, and implement essentially the descent 
operation $I = d I^{(0)}$, $\delta I^{(0)} = d I^{(1)}$, on an auxiliary 
eight-form anomaly polynomial $I$, as required by the Wess-Zumino consistency 
condition.

Interestingly, one can at this point forget about the descent parameter 
$\eta$ associated to the unphysical vertex, keeping in mind that its net 
effect is to implement the descent procedure, and compute directly the anomaly 
polynomial $I$ rather than the anomaly $A$ itself. 
To achieve this out of the remaining correlation, one restricts by hand to the 
eight-form instead of integrating over the six fermionic zero modes which would 
select the six-form, and omits also the integral over the six bosonic zero modes.
We have therefore reduced the computation of the anomaly to the evaluation of a 
supersymmetric partition function in the odd spin-structure in presence of an 
arbitrary gravitational and gauge background. 
World-sheet supersymmetry further implies that this is an index \cite{wit}
which is independent
of the modulus $t$, receiving contribution only from zero energy states of the 
two-dimensional $\sigma$-model. The computation, as anticipated, reduces 
then to the evaluation of the partition function of a supersymmetric quantum 
mechanical model. 
The compact space plays a central role in determining the precise normalization 
factors of the polynomial; moreover, in the cases we will consider in which it 
is realized as an orbifold, it will induce additionally twists in the 
Chan-Paton degrees of freedom \cite{GP}.

The most important outcome of the above reasoning is a simple and general 
prescription for the direct string theory computation of the anomaly inflow 
in arbitrary models, and in particular Type IIB orientifolds. 
By factorizing the annulus, M\"obius strip and Klein bottle inflows, one can 
then deduce the anomalous couplings of all the D-branes, O-planes and fixed-points
in the model. In general, such couplings involve several RR fields arising in all
the twisted sectors, and a generalized inflow mechanism takes places. 
Correspondingly, the anomaly polynomial does not factorize, but rather splits
into a sum of factorized contributions corresponding to a plethora of sub-inflows.


\section{Inflow for K3 orientifolds}

In this section, we shall apply the tools derived in the previous section to
the $D=6$ Type IIB $\Z_N$ orientifolds of \cite{GP,GJ,DP}. In these models, tadpole 
cancellation requires in general both D9-branes and D5-branes, and fixes their number.
In the following, we shall restrict to the special points in the moduli space in
which the models have maximal gauge symmetry. This implies in particular that 
all the D5-branes sit at the fixed-point at the origin of the orbifold. 
For all the details about these models, we refer to \cite{GJ}. 

In this paper, we will focus on the type $A$ $\Z_N$ orientifolds, in the terminology 
of \cite{GJ}. The $D=6$ spectrum consist of gravitational, tensor, hyper and vector
multiplets of $N=(1,0)$ supersymmetry. The untwisted closed string sector yields 
the minimal combination of 1 gravitational, 1 tensor and 2 hyper multiplets for
all $N\neq 2$. For $N=2$ we get 1 gravitational, 1 tensor and 4 hyper multiplets.
The twisted closed string sector gives a varying number of neutral hyper multiplets
and additional tensor multiplets. These closed string spectra are summarized below.

\bigskip
\bigskip
\vbox{
$$\vbox{\offinterlineskip
\hrule height 1.1pt
\halign{&\vrule width 1.1pt#
&\strut\quad#\hfil\quad&
\vrule#
&\strut\quad#\hfil\quad&
\vrule#
&\strut\quad#\hfil\quad&
\vrule width 1.1pt#\cr
height3pt
&\omit&
&\omit&
&\omit&
\cr
&\hfil Model&
&\hfil Neutral Hypers&
&\hfil Tensors&
\cr
height3pt
&\omit&
&\omit&
&\omit&
\cr
\noalign{\hrule height 1.1pt}
height3pt
&\omit&
&\omit&
&\omit&
\cr
&\hfil $\Z_2$&
&\hfil 20&
&\hfil 1&
\cr
height3pt
&\omit&
&\omit&
&\omit&
\cr
\noalign{\hrule}
height3pt
&\omit&
&\omit&
&\omit&
\cr
&\hfil $\Z_3$&
&\hfil 11&
&\hfil 10&
\cr 
height3pt 
&\omit& 
&\omit& 
&\omit&
\cr
\noalign{\hrule }
height3pt 
&\omit& 
&\omit& 
&\omit&
\cr
&\hfil $\Z_4$&
&\hfil 16&
&\hfil 5&
\cr
height3pt 
&\omit& 
&\omit& 
&\omit&
\cr
\noalign{\hrule } 
height3pt 
&\omit& 
&\omit& 
&\omit&
\cr
&\hfil $\Z_6$&
&\hfil 14&
&\hfil 7&
\cr 
height3pt 
&\omit& 
&\omit& 
&\omit&
\cr
}
\hrule height 1.1pt}
$$
}
\bigskip

\noindent
As a consequence of the fact that the total number of neutral hyper
and tensor multiplets is always 21, there are just enough neutral anti-chiral spinors
to combine with the chiral gravitino to give the same gravitational anomaly as 8 
self-dual tensors, as follows from the six-dimensional anomaly cancellation relation 
\be
I_{3/2} - 21 I_{1/2} -  8 I_A = 0\;.
\label{anocanc}
\ee
As a result, the total anomaly from neutral closed string states is 
\be
I_n = (9 - n_T )\,I_A \;,
\ee
where $n_T$ is the number of tensor multiplets in the model.
This anomaly is expected to be canceled by the inflow associated to the 
fixed-plane-fixed-plane interaction encoded in the Klein bottle amplitude.

In the open string sectors, one has vector multiplets as well as charged hyper 
multiplets in various antisymmetric or bi-fundamental representations, as shown
in the following table.

\bigskip
\vbox{
$$\vbox{\offinterlineskip
\hrule height 1.1pt
\halign{&\vrule width 1.1pt#
&\strut\quad#\hfil\quad&
\vrule#
&\strut\quad#\hfil\quad&
\vrule#
&\strut\quad#\hfil\quad&
\vrule width 1.1pt#\cr
height3pt
&\omit&
&\omit&
&\omit&
\cr
&\hfil Model&
&\hfil Gauge Group&
&\hfil Charged Hypermultiplets&
\cr
height3pt
&\omit&
&\omit&
&\omit&
\cr
\noalign{\hrule height 1.1pt}
height3pt
&\omit&
&\omit&
&\omit&
\cr
&\hfil $\Z_2$&
&$\eqalign{
99:\quad U(16)\cr 
55:\quad U(16)\cr
\phantom{95:\quad}}$&
&$\eqalign{
99:\quad 2\times {\bf 120}\cr
55:\quad 2 \times {\bf 120}\cr 
95:\quad ({\bf 16,16})}$&
\cr
height3pt
&\omit&
&\omit&
&\omit&
\cr
\noalign{\hrule}
height3pt
&\omit&
&\omit&
&\omit&
\cr
&\hfil $\Z_3$&
&$\eqalign{99:\quad U(8)\times SO(16)}$&
&$\eqalign{99:\quad  {\bf (28,1)},\;{\bf (8,16)}}$&
\cr 
height3pt 
&\omit& 
&\omit& 
&\omit&
\cr
\noalign{\hrule }
height3pt 
&\omit& 
&\omit& 
&\omit&
\cr
&\hfil $\Z_4$&
& $\eqalign{
99:\quad U(8)\times U(8)\cr
55:\quad U(8)\times U(8)\cr
\phantom{95:\quad}}$&
& $\eqalign{
99:\quad {\bf (28,1)},\;{\bf (1,28)},\;{\bf (8,8)}\cr
55:\quad {\bf (28,1)},\;{\bf (1,28)},\;{\bf (8,8)}\cr
95:\quad ({\bf 8,1;8,1}),\;({\bf 1,8;1,8})}$&
\cr
height3pt 
&\omit& 
&\omit& 
&\omit&
\cr
\noalign{\hrule } 
height3pt 
&\omit& 
&\omit& 
&\omit&
\cr
&\hfil $\Z_6$&
&$\eqalign{ 
99:\quad U(4)\times U(4)\times U(8)\cr 
\phantom{}\cr
55:\quad U(4)\times U(4)\times U(8)\cr
\phantom{}\cr
\phantom{95:\quad}\cr
\phantom{}\cr
\phantom{}}$&
&$\eqalign{
99:\quad \a{\bf (6,1,1)},\;{\bf (1,6,1)}\cr
\a{\bf (4,1,8)},\;{\bf (1,4,8)}\cr
55:\quad \a{\bf (6,1,1)},\;{\bf (1,6,1)}\cr
\a{\bf (4,1,8)},\;{\bf (1,4,8)}\cr
95:\quad \a{\bf (4,1,1;4,1,1)}\cr
\a{\bf (1,4,1;1,4,1)}\cr
\a{\bf (1,1,8;1,1,8)}}$&
\cr 
height3pt 
&\omit& 
&\omit& 
&\omit&
\cr
}
\hrule height 1.1pt}
$$
}
\bigskip

\noindent
In computing the anomalies produced by these fields, it is convenient to decompose
all the representations as tensor products of two fundamental representations (associated 
to the end-points of open strings). Correspondingly, the Chern classes appearing in
the anomaly decompose as products of traces in the fundamental representation.
We shall indicate the latter with $c(F)$ in the following; more precisely, for the 
groups $U(n)$ and $SO(n)$ that appear we define
$$
c(F) = {\rm ch}_{\bf n} (F) = {\rm tr}_{\bf n} [e^{iF/2\pi}] \;.
$$
For $U(n)$, the adjoint, symmetric and antisymmetric representations give
\be
U(n)\;:\;\;\left\{
\begin{array}{l}
\displaystyle{{\rm ch}_{\bf n^2}(F) = c(F)\,c(-F)} \nn \\
\displaystyle{{\rm ch}_{\bf \frac {n(n \pm 1)}2}(F) = \frac 12 \left[
c(F)^2 \pm c(2F)\right]} 
\end{array} \label{u(n)}
\right.\;.
\ee
For $SO(n)$, the adjoint and symmetric representations give similarly
\be
SO(n)\;:\;\;
\displaystyle{{\rm ch}_{\bf \frac {n(n \pm 1)}2}(F) = \frac 12 \left[
c(F)^2 \pm c(2F)\right]} \;.\label{so(n)}
\ee
Using these relations, the total gauge and gravitational anomalies produced by
the charged open string states in the various sectors can be easily computed.
The anomaly polynomials for all the models are reported in the table below. 

\bigskip
\vbox{
$$\vbox{\offinterlineskip
\hrule height 1.1pt
\halign{&\vrule width 1.1pt#
&\strut\quad#\hfil\quad&
\vrule#
&\strut\quad#\hfil\quad&
\vrule width 1.1pt#\cr
height3pt
&\omit&
&\omit&
\cr
&\hfil Model&
&\hfil $I_c$&
\cr
height3pt
&\omit&
&\omit&
\cr
\noalign{\hrule height 1.1pt}
height3pt
&\omit&
&\omit&
\cr
&\hfil $\Z_2$&
&$\eqalign{
\left\{\raisebox{17pt}{}
\left[\raisebox{14pt}{} - c(F_9)^2 + c(F_9)\,c(-F_9)\right]_{99} 
+ \left[\raisebox{14pt}{} - c(F_5)^2 + c(F_5)\,c(-F_5)\right]_{55}\right.\cr
\left.\;\,\,-\left[\raisebox{14pt}{} c(F_5)\,c(F_9)\right]_{95}
+ \left[\raisebox{14pt}{} c(2F_9)\right]_{9} 
+ \left[\raisebox{14pt}{} c(2F_5)\right]_{5}
\raisebox{17pt}{}\right\}I_{1/2}}$&
\cr
height3pt
&\omit&
&\omit&
\cr
\noalign{\hrule}
height3pt
&\omit&
&\omit&
\cr
&\hfil $\Z_3$&
&$\eqalign{\left\{\raisebox{17pt}{}
\displaystyle{\left[\raisebox{14pt}{} - \frac 12 c(F_9^a)^2 + \frac 12 c(F_9)^2
+ c(F_9^a)\,c(-F_9^a) - c(F_9^b)\,c(-F_9^b)\right]_{99}}\right.\cr
\displaystyle{\left.\;\,\,+ \frac 12 \left[\raisebox{14pt}{} c(2F_9^a) - c(2F_9^b)\right]_{9}
\raisebox{17pt}{}\right\}I_{1/2}}}$&
\cr 
height3pt 
&\omit& 
&\omit& 
\cr
\noalign{\hrule }
height3pt 
&\omit& 
&\omit& 
\cr
&\hfil $\Z_4$&
&$\eqalign{
\displaystyle{\left\{\raisebox{17pt}{}\left[\raisebox{14pt}{} 
-\frac 12 \left(c(F_9^a)+c(F_9^b)\right)^2 + c(F_9^a)\,c(-F_9^a) 
+ c(F_9^b)\,c(-F_9^b)\right]_{99}\right.}\cr
\displaystyle{\left.\raisebox{17pt}{}\;\,+\left[\raisebox{14pt}{}
-\frac 12 \left(c(F_5^a)+c(F_5^b)\right)^2 + c(F_5^a)\,c(-F_5^a) 
+ c(F_5^b)\,c(-F_5^b)\right]_{55}\right.} \raisebox{20pt}{} \cr
\displaystyle{\left.\raisebox{17pt}{}\;\,
-\left[\raisebox{14pt}{} c(F_9^a)\,c(F_5^a) + c(F_9^b)\,c(F_5^b)
\right]_{95}\right.} \raisebox{22pt}{} \cr
\displaystyle{\left.\;\,\,+\,\frac 12 
\left[\raisebox{14pt}{} c(2F_9^a) + c(2F_9^b)\right]_9
+\frac 12 \left[\raisebox{14pt}{} c(2F_5^a) + c(2F_5^b)\right]_5 
\raisebox{17pt}{}\right\}I_{1/2}} \raisebox{20pt}{}}$&
\cr
height3pt 
&\omit& 
&\omit&
\cr
\noalign{\hrule } 
height3pt 
&\omit& 
&\omit&
\cr
&\hfil $\Z_6$&
&$\eqalign{
\displaystyle{\left\{\raisebox{17pt}{}\left[\raisebox{14pt}{}
-\frac 12 c(F_9^a)^2 - \frac 12 c(F_9^b)^2 - \left(c(F_9^a) + c(F_9^b)\right)c(F_9^c)
\right.\right.}\cr 
\displaystyle{\left.\raisebox{17pt}{}\left.\;\;\;\,
+ \, c(F_9^a)\,c(-F_9^a) + c(F_9^b)\,c(-F_9^b) + c(F_9^c)\,c(-F_9^c)
\raisebox{14pt}{}\right]_{99}\right.}\cr
\;\;\,\,+\displaystyle{\raisebox{17pt}{}\left[\raisebox{14pt}{}
-\frac 12 c(F_5^a)^2 - \frac 12 c(F_5^b)^2 - \left(c(F_5^a) + c(F_5^b)\right)c(F_5^c)
\right.} \raisebox{20pt}{} \cr \displaystyle{\left.\;\;\;\;\;\;\;\;
+ \, c(F_5^a)\,c(-F_5^a) + c(F_5^b)\,c(-F_5^b) + c(F_5^c)\,c(-F_5^c)
\raisebox{14pt}{} \right]_{55}} \raisebox{20pt}{} \cr
\displaystyle{\left.\raisebox{17pt}{}\;\,-\left[\raisebox{14pt}{}
c(F_9^a)\,c(F_5^a) + c(F_9^b)\,c(F_5^b) + c(F_9^c)\,c(F_5^c)
\raisebox{15pt}{}\right]_{95}\right.} \raisebox{20pt}{} \cr
\displaystyle{\left.\raisebox{17pt}{}\;\,+ 
\frac 12 \left[\raisebox{14pt}{} c(2F_9^a) + c(2F_9^b)\right]_9 
+ \frac 12 \left[\raisebox{14pt}{} c(2F_5^a) + c(2F_5^b)\right]_5 
\raisebox{17pt}{}\right\}I_{1/2}}\raisebox{20pt}{}}$&
\cr 
height3pt 
&\omit& 
&\omit&
\cr
}
\hrule height 1.1pt}
$$
}
\bigskip

\noindent
We have used the fact that in six dimensions only terms with an even number of 
curvatures can appear. For later convenience, we do not expand explicitly the 
anomaly polynomials, but it is understood that only the 8-form component of the 
quoted expressions is relevant. The Latin letters a,b,c,... label the gauge
group factors in the same order as they appear in the previous table. The terms
in the square brackets with index 99, 55 and 95 are expected to be canceled
by inflows associated to the D-brane-D-brane interaction encoded in the annulus 
amplitude in the 99, 55 and 95 sectors. Similarly, the terms in the square brackets 
with index 9 and 5 are expected to be canceled by the inflows associated to the 
D-brane-fixed-planes interaction encoded in the M\"obius strip amplitude in the 
9 and 5 sectors.

Both D-branes (B) and fixed-points (F) are involved in the inflow mechanism. 
The annulus, M\"obius strip and Klein bottle amplitudes in the 
odd spin-structure encode the anomaly inflow arising from the RR magnetic 
interaction between two D-branes (BB), a D-brane and a fixed-point (BF), 
and two fixed-points (FF) respectively. Since there are three types of interaction
among only two kinds of objects, the factorization of these interactions is
non-trivial. The total inflow induced by each surface is obtained by summing 
over all the sectors of the orientifold group the partition function 
of a quantum mechanical model in the (six-dimensional) odd spin-structure.

In the operatorial formalism, the relevant odd spin-structure amplitudes that we
have to compute on the annulus, M\"obius strip and Klein bottle, are given by the 
following partition functions:
\bea
\a\a Z_A = \frac{1}{4N}\sum_{k=0}^{N-1}\,
{\rm Tr}_{R}\,[g^k\,(-1)^F\,e^{-tH(R,F)}] \nn \;,\\ 
\a\a Z_M =\frac{1}{4N}\sum_{k=0}^{N-1}\, 
{\rm Tr}_R\,[\Omega\,g^k\,(-1)^F\,e^{-tH(R,F)}] \label{lc}  \;,\\
\a\a Z_K = \frac{1}{8N}\sum_{k=0}^{N-1}\,\sum_{m=0}^{N-1}
{\rm Tr}^{(m)}_{RR}\,[\Omega\,g^k\,(-1)^{F+\tilde F}\,e^{-tH(R,F)}] \nn \;,
\eea
where a sum over Chan-Paton indices is understood for the case of the annulus and 
M\"obius strip surfaces. In the Klein bottle amplitude, the extra sum includes
all the closed string twisted sectors. The overall factors come form the $\Omega$
and GSO projections.
In the following we will not present the details of the evaluation of the partition
functions of the quantum mechanical models that are reported in \cite{ss,mss}.

\vskip 9pt 
\noindent 
{\bf Annulus}
\vskip 1pt 
\noindent
Consider first the various annulus amplitudes. 
The 99 contribution reads:
\be
Z_A^{99}(R,F_9) = \frac{1}{4N}\,\sum_{k=1}^{N-1}\,N_k\,Z^{B}_k\,Z_k^F\,
{\rm ch}^2(\gamma_k\,F_9)\,\widehat{A}(R) \; ,
\ee
where the Chern class
\be
{\rm ch} (\gamma_k\,F_9) \equiv {\rm tr}\,[\gamma_{k,9}\,
e^{iF_9/2\pi}]
\ee
is defined in the Chan-Paton representation (which is built out of fundamental 
representations of the various factors of the gauge group) and includes the matrix
representation $\gamma_{k,9}$ of the twist $g^k$ induced by the orientifold group action
in the 9 sector. $\widehat{A}(R)$ is the roof genus, and is 
unaffected by the orientifold projection. Finally, $N_k$ and $Z_k^{B,F}$ take into
account the internal partition functions of the four compact bosons and fermions, 
which in the 99 sector have an integer mode expansion and admit zero-energy states. 
The bosons have zero modes $x_0^i\sim g^k\,x_0^i$ in each $k$-twisted 
sector, and one has therefore to sum over all the $k$-fixed-points Fk, whose
number is given by
\be
N_{k} = \left(2\sin \frac {\pi k}{N} \right)^4 \;,\;\; k = 1, \ldots , N-1 \; .
\ee
The fermions have instead zero modes only in the untwisted sector $k=0$, leaving
a vanishing partition function in this sector.
Finally, $Z_k^{B,F}$ represent the partition functions in the $k$-twisted sector of 
the remaining fluctuations of the four internal bosons and fermions
respectively,
and are given by
\be
Z_k^B = \left(2\sin \frac {\pi k}{N} \right)^{-4} \;,\;\;
Z_k^F = \left(2\sin \frac {\pi k}{N} \right)^2 \;,\;\; k=1,...,N-1 \;.
\label{ZFB}
\ee
In the 55 sector the Dirichlet boundary conditions still allow zero-energy states 
for fermions but not for bosons. The corresponding contribution is then:
\be
Z_A^{55}(R,F_5) = \frac{1}{4N}\,\sum_{k=1}^{N-1}\,Z^{F}_k\,
{\rm ch}^2(\gamma_k\,F_5)\,\widehat{A}(R) \; ,
\ee
where now
\be
{\rm ch} (\gamma_k\,F_5) \equiv {\rm tr}\,[\gamma_{k,5}\,
e^{iF_5/2\pi}] \; .
\ee
Finally, in the 95 sector the internal fields satisfy mixed Neumann-Dirichlet 
boundary conditions and have a half-integer mode expansion. This implies that no 
zero-energy states are present in these sectors and correspondingly
\be
Z_A^{95}(R,F_9,F_5) = -\frac{1}{2N}\,\sum_{k=0}^{N-1}\,
{\rm ch} (\gamma_k\,F_9)\,{\rm ch} (\gamma_k\,F_5)\,\widehat{A}(R) \; . \label{95}
\ee
The minus sign in (\ref{95}) is due to the fact that the 95 Ramond
vacuum differs by one unit of fermionic charge with respect to the 99 and 55 
Ramond vacua, and the factor two takes into account the two orientations 95 and 59.
{}From the closed string channel point of view, these expressions encode the 
D9-D9, D5-D5 and D5-D9 magnetic interactions. The $k$-th term in each sum 
corresponds to the exchange of RR forms of the $k$-th twisted sector.

\vskip 9pt 
\noindent 
{\bf M\"obius strip}
\vskip 1pt 
\noindent
The M\"obius strip contribution in the 9 sector reads
\be
I_M^{9}(R,F_9) = -\frac{1}{4N}\,\sum_{k=1}^{N-1}\,N_k\,Z^B_k\,Z_k^F\,
{\rm ch} (\gamma_{\Omega k}^{-1}\gamma_{\Omega k}^T\,2F_9)\,\widehat{A}(R) \;, 
\ee
where again we have taken into account the twist induced by the orientifold group 
action on the Chern class.
The factor two entering in the Chern character is due to the fact that the boundary 
of the M\"obius strip is twice longer than one of the two boundaries of the annulus. 
The analysis for the four internal directions is identical to that presented for the 
99 sector of the annulus and the corresponding contribution is identical and given by
(\ref{ZFB}).
In the 5 sector we get instead
\be
I_M^{5}(R,F_5) = - \frac{1}{4N}\,\sum_{k=0}^{N-1}\,Z^F_{k+N/2}\,
{\rm ch} (\gamma_{\Omega k}^{-1}\gamma_{\Omega k}^T\,2F_5)\,\widehat{A}(R) \;,
\ee
where again the four fermions present zero-energy states, absent for the bosons.
The extra $N/2$-twist is needed to implement Dirichlet boundary conditions
in the M\"obius strip.

In the closed string channel, these expressions correspond to the 
D9-Fk and D5-Fk magnetic interactions. The $k$-th term in each sum involves 
$k$-fixed-points and corresponds to the exchange of RR forms in the $2k$-th 
twisted sector. Correspondingly, the following relations hold (see \cite{GJ})
\be
{\rm ch} (\gamma_{\Omega k}^{-1}\gamma_{\Omega k}^T\,2F) = \pm
{\rm ch} (\gamma_{2 k}\,2F) \;,
\ee
the $+$ and $-$ signs holding in the 9 and 5 sectors respectively.

\vskip 9pt 
\noindent 
{\bf Klein bottle}
\vskip 1pt 
\noindent
In the closed string sector, $\Omega$ exchanges the $m$-th twisted sector with the 
$(N-m)$-th\footnote{Notice that $\Omega$ reported here does {\it not} precisely coincide
with the IIB world-sheet parity operator. See \cite{pol2} for a discussion about this
point.}. This implies that the only sectors which contributes to the Klein bottle
partition function in (\ref{lc}) are the untwisted sector $m=0$ and, for $N$ even, 
the middle sector $m=N/2$. We then get two distinct contributions.
The one coming from the untwisted $m=0$ sector is given by:
\be
I_K(R) = \frac{1}{16N}\,\sum_{k=1}^{N-1}\,N_{k+N/2}\,Z^B_{k+N/2}\,Z^F_k\,
Z_{k+N/2}^F\,\widehat{L}(R)\;,
\label{Kun}
\ee
where we have taken into account that in this sector, both the internal bosons and 
and fermions have zero-energy states and that, analogously to the M\"obius strip case
in the 5 sector, an extra $N/2$-twist is needed to implement the correct
crosscap conditions. Moreover, in this surface left-moving and right-moving fermions
are not simply identified but split into two combinations with periodic and 
anti-periodic zero-energy states respectively \cite{ss}. This explains the various 
factors entering in (\ref{Kun}).

In the $N/2$-twisted sector, the only zero-energy states are the bosonic zero 
modes $x_0^i\sim -x_0^i$. Since in (\ref{lc}) we have also a sum over all the $g^k$-twists 
we get: 
\be
I^\prime_K(R) = -\frac{1}{16N}\,\widehat{L}(R)\,\sum_{k=0}^{N-1}
\,N^{\prime}_k \;,
\label{Kt}
\ee
where $N_k^{\prime}$ represents the number of $N/2$-fixed-points that are also 
$k$-fixed. These numbers are given by $N_0^\prime = N_{N/2}^\prime = 16$, and
$N_k^\prime = N_{N-k}^\prime$ with
\be
N^\prime_k \, = \, \min \, \{N_k,\, N_{k+N/2}\} \;,\;\; k = 1,...,N/2-1 \;.
\ee
The minus sign in (\ref{Kt}) arises from the action of $\Omega$ on the $N/2$-twisted
vacua.

{}From the closed string channel point of view, these two contributions correspond 
respectively to the Fk-Fk and Fk-Fk$\raisebox{1pt}{$\scriptstyle{+}$}\scriptstyle{N/2}$ 
magnetic interactions, and the 
$k$-th term in each sum involves the exchange of RR forms in the $2k$-th twisted 
sector. $\widehat{L}(R)$ indicates the Hirzebruch polynomial, and is crucial for the 
cancellation of anomalies due to (anti)self-dual tensors \cite{ss}.

\vskip 9pt 
\noindent 
{\bf Inflows}
\vskip 1pt 
\noindent
Collecting these results, one finds the following expressions for the various 
inflows:
\bea
\a\a I_{BB}^{99,55}(R,F_{9,5}) = \frac{1}{4N}\,\sum_{k=1}^{N-1}\,
\left(2\sin \frac {\pi k}{N} \right)^2
{\rm ch}^2(\gamma_k\,F_{9,5})\,I_{1/2}(R) \;, \label{IBB9955}\\
\a\a I_{BB}^{95}(R,F_9,F_5) = -\frac{1}{2N}\,\sum_{k=0}^{N-1}\,
{\rm ch}(\gamma_k\,F_9)\,{\rm ch}(\gamma_k\,F_5)\,I_{1/2}(R) \;,\label{IBB95}\\
\a\a I_{BF}^{9}(R,F_9) = -\frac{1}{4N}\,\sum_{k=1}^{N-1}\,
\left(2\sin \frac {\pi k}{N} \right)^2
{\rm ch} (\gamma_{2k}\,2F_9)\,I_{1/2}(R) \;,\label{IBF9}\\
\a\a I_{BF}^{5}(R,F_5) = \frac{1}{4N}\,\sum_{k=0}^{N-1}\,
\left(2\cos \frac {\pi k}{N} \right)^2
{\rm ch} (\gamma_{2k}\,2F_5)\,I_{1/2}(R) \;,\label{IBF5}\\
\a\a I_{FF}(R) =- \frac{1}{2N}\,\sum_{k=0}^{N-1}\,
\left[\left(2\sin \frac {2 \pi k}{N} \right)^2 
- N^\prime_{k}\right] I_{A}(R) \;.\label{IFF}
\eea
In order to compute explicitly these inflows and compare them to the anomalies
in the spectrum, one needs the explicit representation of the $\gamma_k$ matrices.
In a suitable basis, one can choose $\gamma_{k,9} = (\gamma)^k$,
$\gamma_{k,5} = (\gamma^*)^k$, $\gamma^*=\gamma^{-1}$,  
with $\gamma$ given in the following table\footnote{The matrices used here differ from 
those in \cite{GJ} in the fact that there is no relative phase between the matrices
in the 9 and the 5 sectors. The necessity of such a phase is avoided here
by taking explicitly into account the fixed-point degeneracy, achieving a complete 
symmetry between the 9 and the 5 sectors.} 

\bigskip
\vbox{
$$\vbox{\offinterlineskip
\hrule height 1.1pt
\halign{&\vrule width 1.1pt#
&\strut\quad#\hfil\quad&
\vrule#
&\strut\quad#\hfil\quad&
\vrule width 1.1pt#\cr
height3pt
&\omit&
&\omit&
\cr
&\hfil Model&
&\hfil $\gamma$&
\cr
height3pt
&\omit&
&\omit&
\cr
\noalign{\hrule height 1.1pt}
height3pt
&\omit&
&\omit&
\cr
&\hfil $\Z_2$&
&$\eqalign{
{\rm diag}\left(e^{\frac \pi 2 i}\,{\bf I_{16}}, 
e^{-\frac \pi 2 i}\,{\bf I_{\,\overline{\!16\!}\,}}\right)}$&
\cr
height3pt
&\omit&
&\omit&
\cr
\noalign{\hrule}
height3pt
&\omit&
&\omit&
\cr
&\hfil $\Z_3$&
&$\eqalign{
{\rm diag}\left(e^{\frac {2\pi}3 i}\,{\bf I_{8}}^a, 
e^{-\frac {2\pi}3 i}\,{\bf I_{\bar{8}}}^a, 
{\bf I_{16}}^b\right)}$&
\cr 
height3pt 
&\omit&
&\omit&
\cr
\noalign{\hrule }
height3pt 
&\omit& 
&\omit&
\cr
&\hfil $\Z_4$&
&$\eqalign{
{\rm diag}\left(e^{\frac {\pi}4 i}\,{\bf I_{8}}^a, 
e^{-\frac {\pi}4 i}\,{\bf I_{\bar 8}}^a, e^{-\frac {3\pi}4 i}\,{\bf I_{8}}^b, 
e^{\frac {3\pi}4 i}\,{\bf I_{\bar 8}}^b\right)}$&
\cr
height3pt 
&\omit& 
&\omit&
\cr
\noalign{\hrule } 
height3pt 
&\omit& 
&\omit&
\cr
&\hfil $\Z_6$&
&$\eqalign{
{\rm diag}\left(e^{\frac {\pi}6 i}\,{\bf I_{4}}^a, e^{-\frac {\pi}6 i}\,{\bf I_{\bar 4}}^a, 
e^{\frac {5\pi}6 i}\,{\bf I_{4}}^b, e^{-\frac {5\pi}6 i}\,{\bf I_{\bar 4}}^b, 
e^{-\frac {\pi}2 i}\,{\bf I_{8}}^c, e^{\frac {\pi}2 i}\,{\bf I_{\bar 8}}^c\right)}$&
\cr 
height3pt 
&\omit& 
&\omit&
\cr
}
\hrule height 1.1pt}
$$
}

\noindent
Latin letters again refer to the various factors of the gauge group, and 
${\bf I_{\rho}}$ indicates the identity in the representation $\rho$.

It is straightforward but tedious to check that the total inflow, obtained by summing 
all the contributions above, is indeed equal to the total one-loop anomaly $I_n+I_c$, 
for each model. Pictorially, the BB, BF and FF parts of the inflow receive the following 
non-vanishing contributions

\vskip 10pt

\begin{figure}[h]
\epsfxsize = 11.5cm
\vskip 10pt
\hspace{73pt} \epsffile{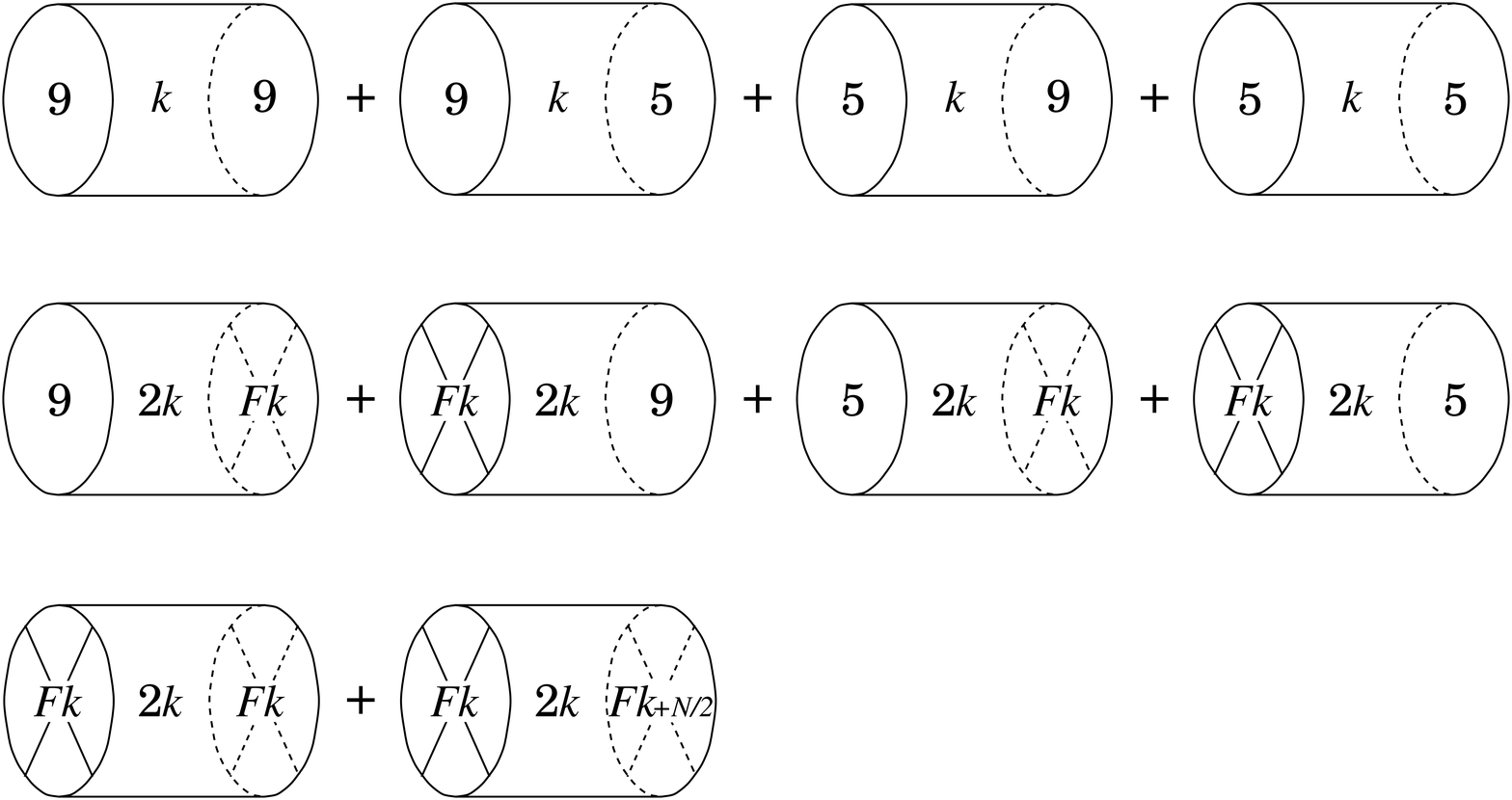}
\vskip -208pt
\bea
I_{BB} \a=\a \displaystyle{\sum_{k=0}^{N-1}} \left[\raisebox{30pt}{} \hspace{333pt} \right]
\;, \nn \\
I_{BF} \a=\a \displaystyle{\sum_{k=0}^{N-1}} \left[\raisebox{30pt}{} \hspace{333pt} \right]
\;, \raisebox{37pt}{} \nn \\
I_{FF} \a=\a \displaystyle{\sum_{k=0}^{N-1}} \left[\raisebox{30pt}{} \hspace{161pt} \right]
\;. \raisebox{37pt}{}
\eea
\vskip -10pt
\end{figure}


\section{Factorization of the couplings}
 
As discussed in previous section, the various tree-level BB, BF and FF inflows 
cancel separately corresponding pieces of the total one-loop anomaly, and are 
interpreted as magnetic interactions. Having computed all the inflows, the 
anomalous couplings to RR fields can be obtained by factorization.
Starting from (\ref{IBB9955})-(\ref{IFF}) and knowing
which (and how) massless fields arise in these orientifold models \cite{GJ}
in all closed string sectors, untwisted and twisted, it is possible to identify 
those responsible for the inflows above. Using the crucial property \cite{mss} 
\be
\sqrt{\widehat{A}(R)}\sqrt{\widehat{L}(R/4)}=\widehat{A}(R/2) \;,
\ee
and making use of the fact that only the 8-form components of all the polynomials 
are relevant to perform some rescalings, it is straightforward to show that the 
inflows (\ref{IBB9955})-(\ref{IFF}) can be indeed consistently factorized.

Recall now that these inflows of anomaly are associated with RR magnetic interactions.
The latter arise in string theory in the odd spin-structure, the insertion of 
$(-1)^F$ acting as Hodge duality on the propagating field \cite{bis}. It is easy to deduce 
from the analysis of \cite{GP,GJ} that the only fields mediating RR magnetic 
interactions are 2-forms in the RR untwisted sector, and 0-forms (scalars) 
magnetically dual to 4-forms, as well as anti-self-dual 2-forms, in the twisted RR sectors.
The content of this kind of fields of the various models is the following.
In the untwisted RR sector, there are always an anti-self-dual and a self-dual 
2-form. They lie in the universal tensor multiplet and in the gravitational 
multiplet respectively, and together they form an unconstrained 2-form $b_{\mu\nu}$.
In the twisted RR sectors, a variable number of scalars and anti-self-dual 2-forms 
occurs, depending on the model. More precisely, one has to distinguish between 
$k$-twisted closed string sectors with $k=N/2$ and $k \neq N/2$. 
At each of the $N_{N/2}=16$ $N/2$-fixed 
points, one gets a scalar $\varphi$ belonging to an extra hypermultiplet. 
Some among these are further identified by the $\Z_N$ projection, and the 
number of independent scalars is 16 for $\Z_2$, 0 for $\Z_3$, 10 for $\Z_4$ and 
6 for $\Z_6$.
Similarly, at each of the $N_{k}$ $k$-fixed-points with $k \neq N/2$, one gets 
a scalar $\phi^{(k)}$, belonging to an extra hypermultiplet, and one anti-self-dual 
2-form $b_{\mu\nu}^{(k)}$, belonging to one extra tensor multiplet. 
Actually, these arise by combining the sectors with twists 
$k$ and $N-k$, which are related by the $\Omega$-projection. 
Again, the $\Z_N$ projection further identifies some of these among each other; the 
net number of them is 0 for $\Z_2$, 9 (from $k=1,2$) for $\Z_3$, 4 (from $k=1,3$) 
for $\Z_4$ and 1+5=6 (from $k=1,5$ and $k=2,4$ respectively) for $\Z_6$.
There is an evident and important distinction arising between the inflows due to 
the exchange of tensor and scalar fields. Tensor fields can have anomalous 
couplings only to 4-forms, and the corresponding contribution to the 8-form anomaly 
polynomial will factorize into the product of two 4-forms. 
Scalar fields can instead have anomalous couplings to 2 and 6-forms, and the 
corresponding contribution to the 8-form anomaly polynomial will factorize into the 
product of a 2-form and a 6-form. This implies that all the inflows induced by exchange 
of scalars and their dual 4-forms involve always the Abelian part of the gauge group. 

Throughout this section, for simplicity, we will
always write at the same time anomalous couplings for forms and their duals,
despite the well known fact that there is no local and covariant Lagrangian in which 
a potential and its dual can appear at the same time. 
More precisely, one should write the couplings
to the dual potential as corrections to the kinetic terms for the field strength
of the original potential, modifying then its Bianchi identity.
In presence of anti-self-dual forms our formulae for the couplings
should be considered as formal expressions that reproduce the right inflow of
anomaly. More precisely, they should be written either in a non-covariant way,
by using the electro-magnetically symmetric action introduced in \cite{sc-se},
or covariantly using the formalism introduced in \cite{pst}.

We now present a general analysis, valid for all the K3 orientifolds considered 
here, allowing to extract in a systematic way the couplings and to emphasize
some general features which are common to all models. In order to keep the analysis 
as general as possible, we find it convenient to keep all the 16 ${\bf Z}_2$-twisted 
scalars and the $N_k$ tensors and scalars arising from the $k$ and $N-k$ twisted
sectors, as independent states. Although the physical propagating states are less 
than that, as mentioned, this will facilitate the general analysis and yet
remain correct. In next section, the anomalous couplings associated to the physical 
scalar fields will be explicitly displayed.

In order to figure out which states are responsible for the various 
inflows, recall that in (\ref{IBB9955})-(\ref{IBB95})
the insertion of $g^k$ acts as a $k$-twist in $\sigma$
for the closed string state exchanged and as a $2k$-twist in $\sigma$
in (\ref{IBF9})-(\ref{IFF}).
It is natural to consider separately the $k=0$, $k=N/2$ closed string sectors
which are special in many respects, and to group the remaining closed string sectors 
into conjugate pairs, adding the contributions of the $k$ and $N-k$ twisted sectors 
which correspond to the same intermediate closed string states.

\vskip 9pt 
\noindent 
{\bf Untwisted closed string exchange}
\vskip 1pt 
\noindent
The non-vanishing contributions to the inflows (\ref{IBB9955})-(\ref{IFF}) involving 
untwisted closed string states of the $k=0$ sector, come form the terms $k=0$ 
in (\ref{IBB95}) and (\ref{IBF5}), $k=N/2$ in (\ref{IBF9}) and $k=0,N/2$ in 
(\ref{IFF}). They can be written as
\bea
\a\a I_{BB}^{(0)95}(R,F_5,F_9) = - Y_{(0)}^9 (F_9,R)\,Y_{(0)}^5 (F_5,R) \;, 
\raisebox{15pt}{} \nn \\
\a\a I_{BF}^{(0)9,5}(R,F_{9,5}) = - Y_{(0)}^{9,5} (F_{9,5},R)\,Y_{(0)} (R) \;,
\raisebox{15pt}{} \nn \\
\a\a I_{FF}^{(0)}(R) = - Y_{(0)}(R)\,Y_{(0)}(R) \;, \raisebox{15pt}{} \label{FIFF0}
\eea
where
\bea
\a\a Y_{(0)}^5 (F_5,R) = \frac 1{\sqrt{2N}}\,
{\rm ch}(F_5)\,\sqrt{\widehat{A}(R)} \;,\nn \\
\a\a Y_{(0)}^9 (F_9,R) = \frac 1{\sqrt{2N}}\,
{\rm ch}(F_9)\,\sqrt{\widehat{A}(R)} \;,\nn \\
\a\a Y_{(0)}(R) = -\frac {32}{\sqrt{2N}}\,\sqrt{\widehat{L}(R/4)} \;.\label{b0}
\eea
\vskip 9pt 
The RR fields involved in the inflow are the untwisted 2-form $b_{\mu\nu}$ and
its magnetic dual $\tilde b_{\mu\nu}$. Form a ten-dimensional point of view, 
$b_{\mu\nu}$ is the trivial dimensional reduction of the RR 2-form, whereas 
$\tilde b_{\mu\nu}$ arises from integrating the RR 6-form on $T^4$. The magnetic
duality of $b_{\mu\nu}$ and $\tilde b_{\mu\nu}$ in six dimensions follows from
the fact that the RR 2 and 6-forms in ten-dimensions have field-strengths 
which are Hodge dual to each other. Form the known form of the anomalous couplings
of D-branes and O-planes in ten-dimensions, we then expect that the combinations
of fields coupling to D5-branes and O5-planes, and D9-branes and O9-planes, 
are simply:
\be
C_{(0)} \, = \, b \;,\;\; \tilde C_{(0)}\,  = \, \tilde b \;.
\label{C0}
\ee
The reason for this is the following. 
D5-branes and O5-planes couple to both the RR 2-form and 6-form in 
ten-dimension. However, the total charge with respect to the latter cancels 
by the requirement of tadpole cancellation, and only the 2-form is relevant, 
producing $b$ in six dimensions. Conversely, since all the internal curvatures 
vanish, the integral over $T^4$ in the anomalous couplings of D9-branes and
O9-planes can act only on the RR 6-form, producing $\tilde b$ in six dimensions.

Using the results of appendix A, it is then straightforward to show that the 
anomalous couplings 
\bea
\a\a S^{(0)}_{D5} = \sqrt{2\pi} \int C_{(0)} \wedge Y_{(0)}^5 (F_5,R) 
\;, \label{AnoD50} \\
\a\a S^{(0)}_{D9} = \sqrt{2\pi} \int \tilde C_{(0)} \wedge Y_{(0)}^9 (F_9,R) 
\;, \label{AnoD90} \\
\a\a S^{(0)}_{O5} = \sqrt{2\pi} \int C_{(0)} \wedge Y_{(0)}(R) 
\;, \label{Ano050} \\
\a\a S^{(0)}_{O9} = \sqrt{2\pi} \int \tilde C_{(0)} \wedge Y_{(0)}(R) 
\;, \label{Ano090}
\eea
reproduce the above inflows. 

\vskip 9pt
\noindent 
{\bf ${\bf N/2}$-twisted closed string exchange}
\vskip 1pt 
\noindent
The non-vanishing contributions to the inflows (\ref{IBB9955})-(\ref{IFF}) involving 
$\Z_2$-twisted closed string states of the $k=N/2$ sector, come from the 
$k=N/2$ terms in (\ref{IBB9955}) and (\ref{IBB95})\footnote{There is also a 
potential contribution from the $k = N/4$ term in the BF and FF inflows 
(\ref{IBF9})-(\ref{IFF}). This exists only for the $\Z_4$ model and happens to cancel 
out.}. They can be written as
\bea
\a\a I_{BB}^{(N/2)99} (R,F_9) = \frac 12\,N_{N/2}
\,Y_{(N/2)}^9(F_9,R)\,Y_{(N/2)}^9(F_9,R) 
\;, \raisebox{15pt}{} \nn \\
\a\a I_{BB}^{(N/2)55} (R,F_5) = \frac 12\,Y_{(N/2)}^5(F_5,R)\,Y_{(N/2)}^5(F_5,R) 
\;, \raisebox{15pt}{} \nn \\
\a\a I_{BB}^{(N/2)95}(R,F_5,F_9) = Y_{(N/2)}^9 (F_9,R)\,Y_{(N/2)}^5 (F_5,R) \;, 
\raisebox{15pt}{} \label{FIBB95N2}
\eea
where $N_{N/2}=16$ and 
\bea
\a\a Y_{(N/2)}^5 (F_5,R) = -\frac 1{\sqrt{2N}}\,2\,{\rm ch}(\gamma_{N/2}\,F_5)
\,\sqrt{\widehat{A}(R)} \;,\nn \\
\a\a Y_{(N/2)}^9 (F_9,R) = \frac 1{\sqrt{2N}}\,2^{-1}\,{\rm ch}(\gamma_{N/2}\,F_9)
\,\sqrt{\widehat{A}(R)} \;.\label{a9N2}
\eea
The RR states responsible for the inflow are the scalars $\varphi^i$ and their
magnetically dual 4-forms $\tilde \varphi^{i}$, with $i=1,...,N_{N/2}$. 
To produce the correct kind of inflows, the anomalous couplings have to involve the 
combinations
\be
C_{(N/2)}^i = \varphi^{i} + \tilde \varphi^{i} \;.
\label{CN2}
\ee
The required anomalous couplings are then found to be
\bea
\a\a S^{(N/2)}_{D5} = \sqrt{2\pi} \int C_{(N/2)}^1 \wedge Y_{(N/2)}^5 (F_5,R) 
\;, \label{AnoD5N2} \\
\a\a S^{(N/2)}_{D9} = \sqrt{2\pi} \sum_{i=1}^{N_{N/2}} 
\int C_{(N/2)}^i \wedge Y_{(N/2)}^9 (F_9,R) \;, \label{AnoD9N2}
\eea
where $i=1$ refers to the fixed-point at the origin, where all the D5-branes sit. 
Further details are reported in appendix A.

\vskip 9pt
\noindent 
{\bf Twisted closed string exchange (${\bf k\neq 0,\,N/2}$)}
\vskip 1pt 
\noindent
In order to extract the inflow associated to $k$ and $N-k$-twisted states, 
we first group the $k$-th and $N-k$-th term of each sum in the inflows 
(\ref{IBB9955})-(\ref{IFF}). The result can be rewritten as a sum over 
$k=1,...,N/2 - 1$ only, with the value $2k = N/2$ excluded for the BF and FF 
inflows, of the following expressions\footnote{It is understood that one has to 
take the integer part of $N/2 - 1$.}
\bea
\a\a I_{BB}^{(k)99}(R,F_9) = \frac{1}{2}\,N_k\,Y^9_{(k)}(F_9,R)\,Y^9_{(k)}(F_9,R) 
\;, \raisebox{16pt}{} \nn \\
\a\a I_{BB}^{(k)55}(R,F_5) = \frac{1}{2}\,Y^5_{(k)}(F_5,R)\,Y^5_{(k)}(F_5,R) 
\;, \raisebox{16pt}{} \nn \\
\a\a I_{BB}^{(k)95}(R,F_5,F_9) = \,Y^9_{(k)}(F_9,R)\,Y^5_{(k)}(F_5,R) 
\;, \raisebox{16pt}{} \nn \\
\a\a I_{BF}^{(2k)9}(R,F_9) = N_k Y^9_{(2k)}(F_9,R) \, Y_{(2k)}(R) 
\;, \raisebox{20pt}{} \nn \\
\a\a I_{BF}^{(2k)5}(R,F_5) = Y^5_{(2k)}(F_5,R) \, Y_{(2k)}(R) 
\;, \raisebox{20pt}{} \nn \\
\a\a I_{FF}^{(2k)}(R) = \frac 12 \left[N_k\,Y_{(2k)}(R)\,Y_{(2k)}(R) 
+ N^\prime_k\,Y_{(2k)}(R)\,Y_{(2k+N)}(R)\right] 
\;, \raisebox{18pt}{} \label{FIFFk} 
\eea
where 
\bea
\a\a Y^5_{(k)}(F_5,R) = -\frac 1{\sqrt{N}}\,\left(2 \sin \frac {\pi k}N\right)\,
{\rm ch}(\gamma_k\,F_5)\,\sqrt{\widehat{A}(R)} \;, \nn  \\
\a\a Y^9_{(k)}(F_9,R) = \frac 1{\sqrt{N}}\,
\left(2 \sin \frac {\pi k}N\right)^{-1}\! 
{\rm ch}(\gamma_k\,F_9)\,\sqrt{\widehat{A}(R)} \;,  \raisebox{21pt}{} \nn \\
\a\a Y_{(2k)}(R) = -\frac {8}{\sqrt{N}}\, \cot \frac {\pi k}N \,
\sqrt{\widehat{L}(R/4)} \;. \raisebox{20pt}{} \label{b2k}
\eea
For each $k=1,...,N/2-1$, the BB inflows in (\ref{FIFFk}) are due the exchange of 
$k$-twisted closed string states, whereas the BF and FF inflows are 
due to the exchange of $2k$-twisted closed string states ($2k \neq N/2$). 
More precisely, recall that the RR states responsible for the inflow in a generic 
$k\neq 0,N/2$ sector are the scalars $\phi_{(k)}^{i_k}$ and their magnetic dual 4-forms 
$\tilde\phi_{(k)}^{i_k}$, as well as the 2-forms $b_{(k)}^{i_k}$ and their dual 
2-forms $\tilde b_{(k)}^{i_k}$, with $i_k=1,...,N_k$. 
As shown in appendix A, in order to produce the correct kind of inflows, the anomalous 
couplings have to involve the combinations 
\be
C_{(k)}^{i_k} = \phi_{(k)}^{i_k} + \tilde \phi_{(k)}^{i_k}
+ \frac 1{\sqrt{2}} (b_{(k)}^{i_k} - \tilde b_{(k)}^{i_k}) \;.
\label{Ck}
\ee
The required anomalous couplings are then given by
\bea
\a\a S^{(k)}_{D5} = \sqrt{2\pi} \int C_{(k)}^1 \wedge Y_{(k)}^5 (F_5,R) 
\;, \label{AnoD5k} \\
\a\a S^{(k)}_{D9} = \sqrt{2\pi} \sum_{i_k=1}^{N_k} 
\int C_{(k)}^{i_k} \wedge Y_{(k)}^9 (F_9,R) \;, \raisebox{25pt}{} 
\label{AnoD9k} \\
\a\a S^{(2k)}_{Fk} = \sqrt{2\pi} \sum_{i_k=1}^{N_k} 
\int C_{(2k)}^{i_k} \wedge Y_{(2k)}(R) \;, \label{AnoFkk}
\eea
for $k=1,...,N/2-1$, with $k\neq N/4$ in (\ref{AnoFkk}). 
Again $i_k=1$ refers to the fixed-point at the origin, where all the D5-branes sit. 
Notice that the fields $C_{(2k)}$ with $2k>N/2$ in (\ref{AnoFkk})
(actually arising only for the $F2$ fixed-point in the ${\bf Z}_6$ model) 
are not independent fields, but rather are defined as 
$C_{(2k)}= \tilde C_{(N - 2k)}$ in terms of those with $2k<N/2$. 
This is at the origin of the second term in the Klein bottle contribution of
(\ref{FIFFk}). Again, some details needed to check that these anomalous couplings 
reproduce the above inflows are reported in appendix A.

\vskip 9pt

Summarizing, the anomalous couplings for D5-branes, D9-branes and Fk-fixed-points, 
in a generic $D=6$ $\Z_N$ orientifold model, are given by equations 
(\ref{AnoD50})-(\ref{Ano090}) in the untwisted sector, (\ref{AnoD5N2}) and 
(\ref{AnoD9N2}) in the $\Z_2$-twisted sector, and (\ref{AnoD5k})-(\ref{AnoFkk}) 
in the other twisted sectors. As usual, it is understood that only the appropriate 
component of the polynomials has to be considered, in this case the 6-form component.
Notice that the inflow completely fixes the coefficients 
in the combinations (\ref{CN2}) and (\ref{Ck}) of the RR tensors, since they are
anti-self dual, but strictly speaking it does not fix completely those
between the RR scalars and dual four-forms. Our choice is just the simplest and
most natural one in which they are taken to be equal. This should be always taken 
into account in the next sections when we write explicitly these couplings. 
Not surprisingly, D5 and D9-branes have RR anomalous couplings in each closed string
sector. O5-planes, corresponding in our language to the $\Z_2$-fixed-points 
$F_{\scriptstyle{N/2}}$, as well as O9-planes, have instead RR anomalous couplings 
only in the untwisted sector. Finally, the other fixed-points $Fk$ have non-vanishing
RR anomalous couplings only in the $2k$-twisted sector. The explicit results
for each of the models we are considering are reported in appendix B.
Notice finally that for $k = 0$ and $k=N/2$, the charges contain only $4p$-forms and 
$4p+2$-forms respectively, whereas for $k \neq 0,N/2$, they contain generically 
$2p$-forms. In the anomalous couplings above, this is compatible with the fact that 
the formal sums $C_{i_k}^{(k)}$ contain only 2-forms for $k=0$, only 0 and 4-forms for 
$k=N/2$, and both 0, 2 and 4-forms for $k \neq 0,N/2$.

\section{Anomaly cancellation}

In this section, we present a more detailed model by model analysis of the inflow 
mechanism. We first identify more precisely which RR fields give rise to the various 
sub-inflows. Applying the general factorization described in last section, we 
then write explicitly all the anomalous couplings for each model, and verify 
that these couplings lead to the correct inflow. This allow to find the 
factorized form of the anomaly and the inflow, exhibiting explicitly the details
of the Green-Schwarz mechanism at work.
As expected, we will see that similarly to what happens in the ${\bf Z}_2$ model 
\cite{americani}, various $U(1)$ factors of the gauge group are spontaneously broken 
by a Higgs mechanism involving the various RR twisted scalars, belonging to 
hypermultiplets. We will discuss this phenomenon in more detail in the next section.
{}For completeness, we report below the analysis of the ${\bf Z}_2$ model,
although a full and detailed discussion of anomaly cancellation in this model 
has been already given in \cite{americani}.

\subsection{$\Z_2$-model}

The RR massless states arising in the model are the following:
\bea
\mbox{Untwisted} &:& \; b_{\mu\nu}\;,\;\; \phi^i \;,\;\; i=1,...,6 \;,\nn \\
\mbox{$\Z_2$-twisted} &:& \; \varphi^j \;,\;\; j=1,...,16 \;.\nn
\eea
Each of the twisted scalars $\varphi^j$ belongs to a six-dimensional $N=1$ 
hypermultiplet. 
By using the general results of last section, it is straightforward to find the 
explicit anomalous couplings for this model, reported in (\ref{AnoZ2}).
Expliciting all the polynomials, the total anomalous term is found to be
$$
\frac {L_{WZ}}{\sqrt{2 \pi}} = b\, X_4^{(5)} \!+ \tilde b\, X_4^{(9)} 
\!+ \varphi^1\, X_6^{(5)} \!+ \tilde \varphi^1\, X_2^{(5)}
\!+ \Big(\frac 14 \sum_{j=1}^{16} \varphi^j\Big)\, X_6^{(9)}
\!+ \Big(\frac 14 \sum_{j=1}^{16} \tilde \varphi^j\Big)\, X_2^{(9)} \;,
$$
where as usual $j=1$ refers to the fixed-point at the origin where the
D5-branes sit and ($\alpha=9,5$)
\bea
\a\a X_4^{(\alpha)} = \frac 1{2(2\pi)^2} \left(\frac 12\,{\rm tr}\,R^2 
- {\rm tr}\,F_\alpha^2 \right) \;, \nn \\
\a\a X_2^{(\alpha)} =\frac {-2}{(2\pi)} \,{\rm tr}\,F_\alpha \;,
\raisebox{20pt}{} \nn \\
\a\a X_6^{(\alpha)} =  \frac {-1}{3(2\pi)^3} 
\left(\frac 1{16}\,{\rm tr}\,R^2 \,{\rm tr}\,F_\alpha
- {\rm tr}\,F_\alpha^3 \right) \;. \raisebox{18pt}{} \nn
\eea
The corresponding inflow is\footnote{The minus sign between the $X_4X_4$ inflows and
the $X_6 X_2$ inflows is due to the conjugation of the gauge curvature in the 
formula for the inflow, see appendix A.}
\be
I\, = \, X_4^{(5)} X_4^{(9)} \!- X_6^{(5)}\Big(X_2^{(5)} \!+ \frac 14 X_2^{(9)}\Big)
- X_6^{(9)}\Big(X_2^{(9)} \!+ \frac 14 X_2^{(5)}\Big) \;.
\label{INZ2}
\ee
This leads to an anomalous variation of the effective action that is 
exactly equal and opposite to the total anomaly $A=2\pi i \int (I_n + I_c)^{(1)}$, 
as can be easily verified. 

\subsection{$\Z_3$-model}

The massless RR states are:
\bea
\mbox{Untwisted} &:& \; b_{\mu\nu} \;,\;\; \phi^i \;,\;\; i=1,...,4 \;, \nn \\
\mbox{$\Z_3$-twisted} &:& \; b^{\prime m}_{\mu\nu} \;,\;\; 
\phi^{\prime m} \;,\;\; m=1,...,9 \;. \nn
\eea
Each of the 9 twisted scalars belong to a hypermultiplet, whereas each of the 9 
antisymmetric tensors belong to a tensor multiplet.
In this case only twisted closed string states participate to the inflow.
The anomalous couplings are again easily found, and are reported in (\ref{AnoZ3}).
Expliciting the polynomials, one then finds
\bea
\frac {L_{WZ}}{\sqrt{2 \pi}} \a=\a \tilde b X_4^{(9)}
\!+ \Big(\frac 13 \sum_{m=1}^9 b^{\prime m}\Big) X_4^{\prime (9)} 
\!- \Big(\frac 13 \sum_{m=1}^{9} \tilde b^{\prime m}\Big) X_4^{\prime (9)} \nn \\ 
\a\;\a \!+ \Big(\frac 13 \sum_{m=1}^{9} \phi^{\prime m} \Big)\, X_6^{\prime (9)}
\!+ \Big(\frac 13 \sum_{m=1}^{9} \tilde \phi^{\prime m} \Big)\, X_2^{\prime (9)} 
\;, 
\eea
where
\bea
\a\a X_4^{(9)} = \frac 1{\sqrt{6}(2\pi)^2} \left[\frac 12\,{\rm tr}\,R^2 
- \left({\rm tr}\,{F_9^a}^2 + \frac 12 \,{\rm tr}\,{F_9^b}^2\right) \right] \;,
\raisebox{15pt}{} \nn \\
\a\a X_4^{\prime (9)} = \frac 1{2\sqrt{2}(2\pi)^2} \left(\frac 14\,{\rm tr}\,R^2 
+ {\rm tr}\,{F_9^a}^2 - {\rm tr}\,{F_9^b}^2 \right) \;,
\raisebox{18pt}{} \nn \\
\a\a X_2^{\prime (9)} = \frac{-\sqrt{3}}{(2\pi)} \,{\rm tr}\,F_9^a \;,
\raisebox{15pt}{} \nn \\
\a\a X_6^{\prime (9)} =  \frac {-1}{2 \sqrt{3}(2\pi)^3} 
\left(\frac 1{16}\,{\rm tr}\,R^2 \,{\rm tr}\,F_9^a
- {\rm tr}\,{F_9^a}^3 \right) \;. 
\raisebox{18pt}{} \nn
\eea
The corresponding inflow is
\be
I = -\, X_4^{\prime (9)} X_4^{\prime (9)} \!- X_6^{\prime (9)} X_2^{\prime (9)} \;,
\label{INZ3}
\ee
and cancels the total anomaly polynomial $I_n+I_c$. 
Notice that the untwisted sector anomalous coupling does not 
contribute to the inflow, but gives nevertheless a non-trivial contribution to 
the Bianchi identity of the antisymmetric tensor of the untwisted sector.

\subsection{$\Z_4$-model}

The RR massless states are:
\bea
\mbox{Untwisted} &:& \; b_{\mu\nu}\;,\;\; \phi^i \;,\;\; i=1,...,4 \;,\nn \\
\mbox{$\Z_4$-twisted} &:& \; b^{\prime m}_{\mu\nu} \;,\;\; 
\phi^{\prime m} \;,\;\; m=1,...,4 \; ,\nn \\
\mbox{$\Z_2$-twisted} &:& \; \varphi^j \;,\;\; j=1,...,10 \; .
\eea
The RR $\Z_4$-twisted scalars and tensors belong respectively to 4 hyper multiplets
and 4 tensor multiplets, whereas the $\Z_2$-twisted scalars belong to 10 hypermultiplets.
Using the generic results found before, one arrives to the anomalous couplings 
(\ref{AnoZ4}), which lead to the following anomalous Lagrangian
\bea
\frac {L_{WZ}}{\sqrt{2 \pi}} \a=\a b\, X_4^{(5)} \!+ \tilde b\, X_4^{(9)} 
\!- b^{\prime 1}\, X_4^{\prime (5)} \!+ \tilde b^{\prime 1}\, X_4^{\prime (5)}
\!+ \Big(\frac 12 \sum_{m=1}^{4} b^{\prime m}\Big) X_4^{\prime (9)} 
\!- \Big(\frac 12 \sum_{m=1}^{4} \tilde b^{\prime m}\Big) X_4^{\prime (9)} \nn \\
\a\;\a \!+ \, \phi^{\prime 1}\, X_6^{\prime (5)} 
\!+ \tilde \phi^{\prime 1}\, X_2^{\prime (5)}
\!+ \Big(\frac 12 \sum_{m=1}^{4} \phi^{\prime m}\Big) X_6^{\prime (9)} 
\!+ \Big(\frac 12 \sum_{m=1}^{4} \tilde \phi^{\prime m}\Big) X_2^{\prime (9)} \\
\a\;\a \!+ \, \varphi^{1} X_6^{(5)} \!+ \tilde \varphi^{1} X_2^{(5)}
\!+ \frac 14\Big(\sum_{j=1}^{4} \varphi^{j} \!+\!
\sqrt{2}\sum_{j=5}^{10}\varphi^{j}\Big) X_6^{(9)}
\!+ \frac 14\Big(\sum_{j=1}^{4} \tilde \varphi^{j} \!+\!
\sqrt{2}\sum_{j=5}^{10}\tilde \varphi^{j}\Big) X_2^{(9)} \;. \raisebox{20pt}{}\nn
\eea
As usual, $j,m=1$ indicates the fixed-point at the origin and
\bea
\a\a X_4^{(\alpha)} = \frac 1{2 \sqrt{2}(2\pi)^2} \left[\frac 12 {\rm tr}\,R^2 
- \left({\rm tr}\,{F_\alpha^a}^2 + {\rm tr}\,{F_\alpha^b}^2 \right)\right] \;, 
\raisebox{16pt}{} \nn \\
\a\a X_4^{\prime (\alpha)} = \frac {-1}{2 \sqrt{2}(2\pi)^2} 
\left({\rm tr}\,{F_\alpha^a}^2 - {\rm tr}\,{F_\alpha^b}^2 \right) \;, 
\raisebox{16pt}{} \nn \\
\a\a X_6^{(\alpha)} = \frac {-1}{3\sqrt{2}(2\pi)^3} \left[\frac 1{16}{\rm tr}\,R^2
\left({\rm tr}\,F_{\alpha}^{a} + {\rm tr}\,F_{\alpha}^{b}\right) 
- \left({\rm tr}\,{F_\alpha^a}^3 + {\rm tr}\,{F_\alpha^b}^3 \right) \right] \;, 
\raisebox{16pt}{} \nn \\
\a\a X_6^{\prime (\alpha)} = \frac {-1}{6(2\pi)^3} \left[\frac 1{16}{\rm tr}\,R^2
\left({\rm tr}\,F_{\alpha}^{a} - {\rm tr}\,F_{\alpha}^{b}\right) 
- \left({\rm tr}\,{F_\alpha^a}^3 - {\rm tr}\,{F_\alpha^b}^3\right) \right] \;, 
\raisebox{16pt}{} \nn \\
\a\a X_2^{(\alpha)} = \frac{-\sqrt{2}}{(2\pi)} \left({\rm tr}\,F_\alpha^{a} 
+ {\rm tr}\,F_\alpha^{b}\right) \;, 
\raisebox{16pt}{} \nn \\
\a\a X_2^{\prime(\alpha)} = \frac {-1}{(2\pi)} \left({\rm tr}\,F_\alpha^{a} 
- {\rm tr}\,F_\alpha^{b}\right) \;.
\raisebox{18pt}{}
\eea
The corresponding inflow is
\bea
I \a=\a X_4^{(5)} X_4^{(9)} 
\!- X_4^{\prime(5)} \Big(X_4^{\prime(5)} \!- \frac 12 X_4^{\prime(9)}\Big)
- X_4^{\prime(9)} \Big(X_4^{\prime(9)} \!- \frac 12 X_4^{\prime(5)}\Big) \nn \\
\a\;\a -\, X_6^{\prime(5)} \Big(X_2^{\prime(5)} \!+ \frac 12 X_2^{\prime(9)}\Big)
- X_6^{\prime(9)} \Big(X_2^{\prime(9)} \!+ \frac 12 X_2^{\prime(5)}\Big) \nn \\
\a\;\a -\, X_6^{(5)} \Big(X_2^{(5)} \!+ \frac 14 X_2^{(9)}\Big)
- X_6^{(9)} \Big(X_2^{(9)} \!+ \frac 14 X_2^{(5)}\Big) \;,
\label{INZ4}
\eea
that is equal and opposite to the total anomaly polynomial.

The form of the $U(1)$ combinations coupling to the various scalars can be checked
as follows. Whenever a $k$-twisted scalar field couples to a $U(1)$ gauge field, 
the corresponding two-point function on the disk, with the gauge field vertex 
operator on the boundary and the closed string scalar vertex operator in its interior,
has to be non-vanishing. Denoting with $\lambda$ the Chan-Paton wave function 
associated to the two $U(1)$ gauge fields, due to the $k$-twist, the amplitude in 
question will be proportional to ${\rm tr}\,(\lambda\gamma_k)$. 
It is easy to see that with our choice of basis, the Chan-Paton wave functions for 
the $U(1)$ combinations $F_\alpha^a \mp F_\alpha^b$ are 
$\lambda^{1,2}={\rm diag}\,({\bf I_8},-{\bf I_8},\mp{\bf I_8},\pm{\bf I_8})$, 
in both the $\alpha=9,5$ sectors. It is then clear that
${\rm tr}\,(\lambda^i \gamma_k) \sim \delta^i_k$, in agreement with the fact that
the inflows due to the $k=1$ and $k=2$ twisted scalars do not mix.

\subsection{$\Z_6$-model}

The RR massless states are:
\bea
\mbox{Untwisted} &:& \; b_{\mu\nu}\;,\;\; \phi^i \;,\;\; i=1,...,4 \;,\nn \\
\mbox{$\Z_6$-twisted} &:& \; b^{\prime}_{\mu\nu} \;,\;\; 
\phi^{\prime} \;, \nn \\
\mbox{$\Z_3$-twisted} &:& \; b^{\prime\prime m}_{\mu\nu} \;,\;\; 
\phi^{\prime\prime m} \;,\;\; m=1,...,5 \; , \nn \\
\mbox{$\Z_2$-twisted} &:& \; \varphi^j \;,\;\; j=1,...,6 \; . 
\eea
The RR $\Z_6$-twisted scalars and tensors belong to 1 hyper and 1 tensor 
multiplet, the $\Z_3$-twisted ones to 5 hyper and 5 tensor multiplets, and 
the $\Z_2$-twisted scalars to 6 hypermultiplets. 
The various anomalous couplings for this model are reported in (\ref{AnoZ6}),
and lead to
\bea
\frac {L_{WZ}}{\sqrt{2 \pi}} \a=\a b\, X_4^{(5)} \!+ \tilde b\, X_4^{(9)} 
\!+ b^{\prime} \Big(X_4^{\prime (9)} \!- X_4^{\prime (5)}\Big)  
- \tilde b^{\prime} \Big(X_4^{\prime (9)} \!- X_4^{\prime (5)}\Big) \nn \\
\a\;\a \!+ \,\phi^{\prime} \Big(X_6^{\prime (5)} \!+ X_6^{\prime (9)}\Big) 
+ \tilde \phi^{\prime} \Big(X_2^{\prime (5)} \!+ X_2^{\prime (9)}\Big) 
- b^{\prime\prime 1} X_4^{\prime\prime (5)} \!+ \tilde b^{\prime\prime 1} 
X_4^{\prime\prime (5)} \nn \raisebox{16pt}{} \\
\a\;\a \!+ \frac 13 \Big(b^{\prime\prime 1} \!+\! \sqrt{2} \!\sum_{m=2}^{5} 
b^{\prime\prime m}\Big) X_4^{\prime\prime (9)} 
\!- \frac 13 \Big(\tilde b^{\prime\prime 1} \!+\! \sqrt{2} \!\sum_{m=2}^{5} 
\tilde b^{\prime\prime m}\Big) X_4^{\prime\prime (9)} 
\!+ \phi^{\prime\prime 1} X_6^{\prime\prime (5)} 
\!+ \tilde \phi^{\prime\prime 1} X_2^{\prime\prime (5)} \nn \raisebox{24pt}{} \\
\a\;\a \!+ \frac 13 \Big(\phi^{\prime\prime 1} \!+\! \sqrt{2} \!\sum_{m=2}^{5} 
\phi^{\prime\prime m}\Big) X_6^{\prime\prime (9)} 
\!+ \frac 13 \Big(\tilde \phi^{\prime\prime 1} \!+\! \sqrt{2} \!\sum_{m=2}^{5} 
\tilde \phi^{\prime\prime m}\Big) X_2^{\prime\prime (9)}
\!+ \varphi^{1} X_6^{(5)} \!+ \tilde \varphi^{1} X_2^{(5)} \nn \\ 
\a\;\a \!+ \frac 14 \Big(\varphi^{1} \!+\! 
\sqrt{3} \sum_{j=2}^{6} \varphi^{j}\Big)X_6^{(9)} 
\!+ \frac 14\Big(\tilde \varphi^{1} \!+\! \sqrt{3} \sum_{j=2}^{6} 
\tilde \varphi^{j}\Big)X_2^{(9)} \;,
\label{Z6wz}
\eea
where
\bea
\a\a X_4^{(\alpha)} = \frac 1{2 \sqrt{3}(2\pi)^2} \left[\frac 12 {\rm tr}\,R^2 
- \left({\rm tr}\,{F_\alpha^a}^2 + {\rm tr}\,{F_\alpha^b}^2 
+ {\rm tr}\,{F_\alpha^c}^2 \right)\right] \;, 
\raisebox{16pt}{} \nn \\
\a\a X_4^{\prime (\alpha)} = \frac {-1}{4(2\pi)^2}
\left({\rm tr}\,{F_\alpha^a}^2 - {\rm tr}\,{F_\alpha^b}^2 \right) \;, 
\raisebox{16pt}{} \nn \\
\a\a X_4^{\prime\prime (\alpha)} =  \frac {-1}{4(2\pi)^2} \left[\frac 14 {\rm tr}\,R^2
+ \left({\rm tr}\,{F_\alpha^a}^2 + {\rm tr}\,{F_\alpha^b}^2 
-2 \,{\rm tr}\,{F_\alpha^b}^2\right)\right] \;, 
\raisebox{18pt}{} \nn \\
\a\a X_6^{(\alpha)} = \frac {-1}{3\sqrt{3}(2\pi)^3} \left[\frac 1{16}{\rm tr}\,R^2
\left({\rm tr}\,F_{\alpha}^{a} + {\rm tr}\,F_{\alpha}^{b} 
+ {\rm tr}\,F_{\alpha}^{c}\right) 
- \left({\rm tr}\,{F_\alpha^a}^3 + {\rm tr}\,{F_\alpha^b}^3 
+ {\rm tr}\,{F_{\alpha}^c}^3\right) \right] \;, 
\raisebox{16pt}{} \nn \\
\a\a X_6^{\prime (\alpha)} = \frac {-1}{6\sqrt{6}(2\pi)^3} \left[\frac 1{16}{\rm tr}\,R^2
\left({\rm tr}\,F_{\alpha}^{a} + {\rm tr}\,F_{\alpha}^{b} 
- 2 \, {\rm tr}\,F_{\alpha}^{c}\right) 
- \left({\rm tr}\,{F_\alpha^a}^3 + {\rm tr}\,{F_\alpha^b}^3 
- 2 \, {\rm tr}\,{F_{\alpha}^c}^3\right) \right] \;, 
\raisebox{16pt}{} \hspace{-10pt} \nn \\
\a\a X_6^{\prime\prime (\alpha)} = \frac {-1}{3\sqrt{6}(2\pi)^3} 
\left[\frac 1{16}{\rm tr}\,R^2 \left({\rm tr}\,F_{\alpha}^{a} 
- {\rm tr}\,F_{\alpha}^{b}\right) - \left({\rm tr}\,{F_\alpha^a}^3 
- {\rm tr}\,{F_\alpha^b}^3\right) \right] \;, 
\raisebox{15pt}{} \nn \\
\a\a X_2^{(\alpha)} = \frac {-2}{\sqrt{3}(2\pi)} \left({\rm tr}\,F_\alpha^{a} 
+ {\rm tr}\,F_\alpha^{b} + {\rm tr}\,F_\alpha^{c}\right) \;, 
\raisebox{16pt}{} \nn \\
\a\a X_2^{\prime (\alpha)} =  \frac {-1}{\sqrt{6}(2\pi)}
\left({\rm tr}\,F_\alpha^{a} + {\rm tr}\,F_\alpha^{b} - 2 \,{\rm tr}\,F_\alpha^{c}\right) \;,
\raisebox{16pt}{} \nn \\
\a\a X_2^{\prime\prime (\alpha)} = \frac {-3}{\sqrt{6}(2\pi)}
\left({\rm tr}\,F_\alpha^{a} - {\rm tr}\,F_\alpha^{b}\right) \;.
\raisebox{16pt}{}
\eea
The corresponding inflow is
\bea
I \a=\a X_4^{(5)} X_4^{(9)} \!- \Big(X_4^{\prime (5)} \!- X_4^{\prime (9)}\Big) 
\Big(X_4^{\prime (5)} \!- X_4^{\prime (9)}\Big) 
- \Big(X_6^{\prime(5)} \!+ X_6^{\prime(9)}\Big) 
\Big(X_2^{\prime(5)} \!+ X_2^{\prime(9)}\Big) \nn \\
\a\;\a -\, X_4^{\prime\prime(5)} \Big(X_4^{\prime\prime(5)} 
\!- \frac 13 X_4^{\prime\prime(9)}\Big)
- X_4^{\prime\prime(9)} \Big(X_4^{\prime\prime(9)} 
\!- \frac 13 X_4^{\prime\prime(5)}\Big) \nn \\
\a\;\a -\, X_6^{\prime\prime(5)} \Big(X_2^{\prime\prime(5)} 
\!+ \frac 13 X_2^{\prime\prime(9)}\Big)
- X_6^{\prime\prime(9)} \Big(X_2^{\prime\prime(9)} 
\!+ \frac 13 X_2^{\prime\prime(5)}\Big) \nn \\
\a\;\a -\, X_6^{(5)} \Big(X_2^{(5)} \!+ \frac 14 X_2^{(9)}\Big)
- X_6^{(9)} \Big(X_2^{(9)} \!+ \frac 14 X_2^{(5)}\Big) \;,
\label{INZ6}
\eea
and is equal and opposite to the total anomaly polynomial.

Again, the form of the $U(1)$ combinations coupled to the various scalars can be 
checked through a disk computation, which in the $k$-twisted sector is again
proportional to ${\rm tr}\,(\lambda\gamma_k)$. The Chan-Paton wave functions for 
the three $U(1)$ combinations $F_\alpha^a + F_\alpha^b -2 \,F_\alpha^c$, 
$F_\alpha^a - F_\alpha^b$ and $F_\alpha^a + F_\alpha^b + F_\alpha^c$ are
$\lambda^{1}={\rm diag}\,({\bf I_4},-{\bf I_4}, {\bf I_4},-{\bf I_4},
-{\bf I_8},{\bf I_8})$, $\lambda^{2}={\rm diag}\,({\bf I_4},-{\bf I_4},
-{\bf I_4},{\bf I_4},0\, {\bf I_8},0\, {\bf I_8})$ and 
$\lambda^{3}={\rm diag}\,({\bf I_4},-{\bf I_4}, 
{\bf I_4},-{\bf I_4},1/2 \,{\bf I_8},-1/2 \,{\bf I_8})$.
Then, one finds as before ${\rm tr}\,(\lambda^i \gamma_k) \sim \delta^i_k$, again 
in agreement with the fact that the inflows due to the $k=1$, $k=2$ and $k=3$ twisted 
scalars do not mix.


\section{Field theory analysis}

In this section, we analyze some aspects of the low-energy effective action of the 
orientifold models discussed here. In particular, we focus on some interesting 
implications of the anomaly cancelling couplings found in section five and other terms
related to them by $N=1$ $D=6$ supersymmetry.

\subsection{Anomalous $U(1)$'s}

As already shown in detail for the ${\bf Z}_2$ case in \cite{americani},
a generalized Green-Schwarz mechanism involving scalars induces an
Higgs mechanism breaking various $U(1)$ factors of the gauge group.
As expected, this feature is common to all IIB orientifold models
discussed here. Differently from the ${\bf Z}_2$ case, however, one has in general 
to carefully establish whether the RR twisted scalars belong to hyper or tensor 
multiplets. Luckily, the answer can be derived from \cite{abd}, where it was shown 
that the scalars belonging to the extra tensor multiplets are actually twisted NSNS 
scalars\footnote{We thank A. Sagnotti for very useful discussions about this point.}. 
This implies that all the RR twisted scalars involved in the inflow mechanism 
analyzed in section five belong 
to neutral hypermultiplets, like in the ${\bf Z}_2$ case. 
The analysis is then identical to that of \cite{americani}. After an integration by parts 
and a duality transformation, the anomalous couplings to the various four-forms reported in 
last section enter as corrections to the kinetic terms of the corresponding dual scalars.
For these fields strength to be gauge invariant, the scalars have then to transform 
anomalously under the corresponding $U(1)$ gauge transformation. 
In order to write explicit expression, it is convenient to use the linear combinations
of scalar fields entering the inflows as new independent fields $\lambda^I$. 
By doing so, the combinations $A^I$ of $U(1)$ gauge fields entering as shifts in the 
field-strengths
\be
H^I = d \lambda^I - \frac{A^I}{2\pi}
\ee
are directly related to the combinations $X_2^I$ of 2-forms in the corresponding factor of 
the inflows derived in section five, that is (\ref{INZ2}), (\ref{INZ3}), (\ref{INZ4}) and 
(\ref{INZ6}). The precise relation is $X_2^{I} = -1/\pi d A^I$, so that 
the gauge field $A^I$ is essentially the first descent of $X_2$: 
$A^I = - \pi X_2^{I(0)}$. Also the anomalous 
transformations are easily deduced: under the gauge transformation 
$\delta A^I = d \epsilon^I$, the scalars transform as $\delta \lambda^I = \epsilon^I/2\pi$.

Using the general results and the notation of section four, the anomalous interaction 
leading to the shift in the field strengths of the twisted scalars $\phi^{(k)}$ can be 
written in a very compact form. One finds
$$
S^{(k)}_{SC} = \sqrt{\frac {2}{\pi N}} \sin \frac {\pi k}N \int \! d^6x 
\left\{\partial_\mu \phi^{(k)1} {\rm tr}\,[\gamma_k \, A^\mu_5] +
\Big(\frac 1{\sqrt{N_k}} \sum_{i_k=1}^{N_k} \partial_\mu \phi^{(k)i_k}\Big) 
{\rm tr}\,[\gamma_k \, A^\mu_9] \right\} \;.
$$
In the following, we report explicit expressions for each model in terms of the physical 
scalars\footnote{Recall that some of the $\phi^{(k)}$'s are identified. See Appendix B.}.

\vskip 9pt
\noindent 
{\bf $\Z_2$-model}
\vskip 1pt 
\noindent
In the ${\bf Z}_2$ case, the combinations of scalars participating to the inflow and the 
corresponding $U(1)$ gauge fields are given by
$$
\begin{array}{lll}
\lambda^1 = \varphi^1 \a\;:\;\a
\ds{A^1 =  A_9 + \frac 14 A_5 } \nn \;, \\
\ds{\lambda^2 = \frac 14 \sum_{j=1}^{16} \varphi^j} \a\;:\;\a
\ds{A^2 = A_5 + \frac 14 A_9 } \;. \nn
\end{array}
$$
The two scalars $\lambda^{1,2}$ are eaten by the two $U(1)$ gauge fields $A^{1,2}$ through a 
Higgs mechanism, and the latter become massive. Since supersymmetry remains unbroken, the 
whole hypermultiplets containing $\lambda^{1,2}$ must become massive. This is indeed what 
happens since the inhomogeneous $U(1)$ gauge transformations of the scalars $\lambda^{1,2}$ 
turn some D-terms into mass terms for the remaining scalars in the hypermultiplet 
\cite{damo,americani} and, most likely, the two Weyl fermions of opposite chiralities, 
belonging to the vector and hyper multiplet, combine into a single massive Dirac fermion. 
Since in this Higgsing the only states that disappear from the massless spectrum and 
contribute 
to gauge and gravitational anomalies are the two massless chiral fermions, of opposite 
chiralities, the model remains anomaly-free, with unbroken $(SU(16))^2$ gauge group.

\vskip 9pt
\noindent 
{\bf $\Z_3$-model}
\vskip 1pt 
\noindent
In the ${\bf Z}_3$ case, there is only one combination of scalars and one $U(1)$ gauge 
field, given by
$$
\lambda = \frac 13 \sum_{m=1}^{9} \phi^{\prime m} \;:\;
A = \frac {\sqrt{3}}{2} A_9^a \;. \nn
$$
By the same mechanism as before, the unbroken gauge group is found to be 
$SO(16)\times SU(8)$.

\vskip 9pt
\noindent 
{\bf $\Z_4$-model}
\vskip 1pt 
\noindent
The ${\bf Z}_4$ model presents four different combinations of scalars and $U(1)$ gauge 
fields, given by
$$
\begin{array}{lll}
\lambda^1 = \phi^{\prime 1} \a\;:\;\a
\ds{A^1 = \frac 1{2} \left[\left(A_5^a - A_5^b\right) 
+ \frac 12 \left(A_9^a - A_9^b \right)\right]} \nn \;, \\
\ds{\lambda^2 = \frac 12 \sum_{m=1}^4 \phi^{\prime m}} \a\;:\;\a
\ds{A^2 = \frac 1{2} \left[\left(A_9^a - A_9^b\right) 
+ \frac 12 \left(A_5^a - A_5^b\right)\right]} \nn \;, \\
\lambda^3 = \varphi^{1} \a\;:\;\a
\ds{A^3 = \frac 1{\sqrt{2}} \left[\left(A_5^a + A_5^b\right) 
+ \frac 14 \left(A_9^a + A_9^b\right)\right]} \nn \;, \\
\lambda^4 = \ds{\frac 14 \Big(\sum_{j=1}^{4} \varphi^{j} \!+\!
\sqrt{2} \! \sum_{j=5}^{10}\varphi^{j}\Big)} \a\;:\;\a
\ds{A^4 = \frac 1{\sqrt{2}} \left[\left(A_9^a + A_9^b\right) 
+ \frac 14 \left(A_5^a + A_5^b\right)\right]} \nn \;.
\end{array}
$$
All the four $U(1)$ factors are then spontaneously broken, leaving as unbroken gauge group 
$(SU(8) \times SU(8))^2$.

\vskip 9pt
\noindent 
{\bf $\Z_6$-model}
\vskip 1pt 
\noindent
The ${\bf Z}_6$ model presents the new feature that a $U(1)$ factor is left unbroken. 
Indeed, there are only five independent combinations of scalars involved in the Higgs 
mechanism, corresponding to five combinations of the six $U(1)$ fields:
$$
\begin{array}{lll}
\lambda^1 = \phi^{\prime} \a\;:\;\a
\ds{A^1 = \frac 1{2\sqrt{6}} \left[\left(A_5^a + A_5^b - 2 \,A_5^c \right) 
+ \left(A_9^a + A_9^b - 2 \,A_9^c\right) \right]} \nn \;, \\
\lambda^2 = \phi^{\prime\prime 1} \a\;:\;\a 
\ds{A^2 =  \frac 3{2\sqrt{6}} \left[\left(A_5^a - A_5^b\right) 
+ \frac 13 \left(A_9^a - A_9^b\right)\right]} \nn \;, \\
\ds{\lambda^3 = \frac 13 \Big(\phi^{\prime\prime 1} \!+\! 
\sqrt{2} \!\sum_{m=2}^5 \phi^{\prime m} \Big)} \a\;:\;\a
\ds{A^3 = \frac 3{2\sqrt{6}} \left[\left(A_9^a - A_9^b\right) 
+ \frac 13 \left(A_5^a - A_5^b\right)\right]} \nn \;, \\
\lambda^4 = \varphi^{1} \a\;:\;\a
\ds{A^4 = \frac 1{\sqrt{3}} \left[\left(A_5^a + A_5^b + A_5^c\right) 
+ \frac 14 \left(A_9^a + A_9^b + A_5^c\right) \right]} \nn \;, \\
\ds{\lambda^5 = \frac 14 \Big(\sum_{j=1}^{4} \varphi^{j} \!+\!
\sqrt{3} \sum_{j=2}^{6}\varphi^{j}\Big)} \a\;:\;\a 
\ds{A^5 = \frac 1{\sqrt{3}} \left[\left(A_9^a + A_9^b + A_5^c\right) 
+ \frac 14 \left(A_5^a + A_5^b + A_5^c\right) \right]} \nn \;.
\end{array}
$$
The resulting unbroken gauge group is $(SU(4) \times SU(4) \times SU(8))^2 \times U(1)$.
It is straightforward to determine how the surviving $U(1)$ factor is embedded in the 
original gauge group. Indeed, the corresponding gauge field is the linear combination 
which is orthogonal to $A^{1-5}$, and is found to be similar to $A^1$, but with a relative
minus sign between 5 and 9 factors.

\vskip 5pt 

A similar analysis can be performed in the more general case where we have
arbitrary configurations for the D5 branes and non-vanishing Wilson lines in
the nine branes gauge groups. Although we do not report this more general analysis, it 
is clear that in any configuration the maximum number of $U(1)$'s that can be Higgsed
cannot exceed the number of RR twisted closed string scalars present in each model.

\subsection{Gauge couplings}

Let us now turn our attention to the Chern-Simons couplings of the tensor fields.
The most remarkable property of these terms is that their structure is related by 
supersymmetry to the gauge couplings constants $1/g^2$ of the theory \cite{sagGS}. 
Whenever some of them vanish or become negative, tensionless strings appear and the 
model undergoes a phase transition \cite{dmw,sw,dlpo}. It is then extremely useful to 
analyze in detail anomaly cancellation, being a useful tool to address such questions.

In order to be more concrete, we briefly review here some general properties of $N=1$ $D=6$
supergravity coupled to $n_T$ tensor multiplets and vector multiplets of some gauge group.
For details we refer the reader to \cite{frs} and references therein.
The coupling to the hypermultiplet sector does not play any role in the discussion that
will follow and it will be omitted\footnote{See \cite{ns} for an analysis of the hypermultiplet
sector.}. The kinetic terms for the bosonic fields of the theory, including the 
Chern-Simons couplings for the two-forms, are \cite{frs}
\bea
e^{-1}{\cal L}_{bos.}^{kin.} \a=\a  - \frac 14 R + \frac 1{12} G_{rs}\,H_{\mu\nu\rho}^r\, 
H^{s,\mu\nu\rho}  - \frac 14 (\partial_{\mu}v^r)(\partial^{\mu}v_r) \nn \\ 
\a\;\a - \frac 12 v_r\, c^r_z \, {\rm tr}_z\, F_{\mu\nu}F^{\mu\nu}
- 6\,e^{-1}\, c_{rz}\, b^r \wedge {\rm tr}_z \, F^2
\;. \label{Lsugra}
\eea
The physical scalars $v_r$, $r=0,..., n_T$, parametrize the coset space 
$SO(1,n_T)/SO(n_T)$, whereas the non-propagating scalars $x_r^M$, $M=1,...,n_T$,
can be gauged away by fixing the local $SO(n_T)$ symmetry; together, $v_r$ and $x_r^M$
form an $SO(1,n_T)$ matrix. Moreover, $e$ is the determinant of the sechsbein, and 
$G_{rs}=v_rv_s+x_r^Mx_s^M$. Finally, $H^r=db^r-c^r_z w_3^z$ are the modified 
field-strengths of the tensors, shifted by the Chern-Simons three-forms $w_3^z$ 
defined as descents of ${\rm tr}_z \, F^2$: $d w_3^z={\rm tr}_z \, F^2$; 
the $c^r_z$ are constants, and $z$ labels different factors of the gauge group.
The crucial observation of \cite{sagGS}
is that the same coefficients $c^r_z$ appear in the Bianchi identities and couplings
for $b^r$ and in the gauge couplings. Among the $n_T+1$ field strengths $H^r$,
$H^{(+)}=v_r\,H^r$ is self-dual, while $K^{M(-)}=x^M_r\, H^r$ are anti-self-dual.
The Lagrangian (\ref{Lsugra}) presents a manifest global $SO(1,n_T)$ symmetry rotating
the tensor fields, the scalars and the constants $c^r_z$. In writing a local covariant 
Lagrangian as in (\ref{Lsugra}), it is understood that the (anti)self-duality constraints
for the field strengths must be imposed only after varying the Lagrangian, at the level
of equations of motion. Alternatively, following the work of \cite{pst},
one could add an auxiliary scalar field and additional terms to the model that take into
account the (anti)self-duality conditions. We will not do that, since this will be 
irrelevant for our considerations, but notice that such Lagrangian, including all 
four-fermion terms, already appeared in \cite{rs}. 

In order to better understand how to apply these 
general considerations to the IIB orientifold models analyzed here, it is useful
to consider first a model with $n_T=1$ and arbitrary gauge group.
In this case, the $SO(1,1)$ matrix is simply parametrized by $v_0=\cosh \phi=x_1^1$,
$v_1=\sinh \phi=x_0^1$. We can also combine the self-dual and anti-self-dual tensors
to form an unconstrained two-form field. Choosing the simple combination
$b=1/2(b^0-b^1)$, the Lagrangian (\ref{Lsugra}) becomes
\bea
e^{-1}\, {\cal L}_{bos.}^{kin.} \a=\a-\frac 14 R + \frac 16 e^{-2\phi}\,
H_{\mu\nu\rho}\, 
H^{\mu\nu\rho} - \frac 14 (\partial_{\mu}v^r)(\partial^{\mu}v_r) \nn \\
\a\;\a - \frac 14 (e^{\phi}\, c_z +  e^{-\phi}\,\tilde{c}_z)
\, {\rm tr}_z F_{\mu\nu}F^{\mu\nu} -  6\,e^{-1}\, c_{z}\, b \wedge {\rm tr}_z \, F^2
\; , \label{Lsugra1}
\eea
where $c_z=c^0_z+c^1_z$, $\tilde c_z=c^0_z-c^1_z$ and $H=db-\tilde c_z/2 \, w_3^z$.
We dropped from (\ref{Lsugra1}) the other linear combination $\tilde b=1/2(b^0+b^1)$
since its dynamics is completely determined from that of $b$.
Let us connect this Lagrangian with the one expected from type I compactified on K3.
In presence of D5-branes, the ten dimensional 
kinetic terms for $b$ and the $F_{9,5}$ field strengths are schematically, in the string frame:
\be
e_S^{-1}\,{\cal L}^{(10)}_{kin.}\sim e^{-2D_{10}}\, R_S \, + H^2 \, + e^{-D_{10}}\,
{\rm tr} F_9^2 \, + \, e^{-D_{10}} \, \delta^{(4)}(y) \, {\rm tr} F_5^2 \; ,
\ee
with $D_{10}$ the ten dimensional dilaton. Compactifying on $D=6$ and going to the
Einstein frame, one gets
\be
e_E^{-1}\,{\cal L}^{(6)}_{kin.}\sim  R_E \, + \, V_4 \, H^2 \, + \, V_4^{1/2} \,
{\rm tr} F_9^2 \, + \, V_4^{-1/2} \, {\rm tr} F_5^2 \; , \label{Lstr}
\ee
where $V_4$ is the volume of $K3$. 
By comparing (\ref{Lstr}) with (\ref{Lsugra1}), we see
that $V_4 = e^{-2 \langle \phi \rangle}$, $\phi$ being the scalar in the tensor multiplet.
Moreover the comparison requires also that $c_9=\tilde c_5=0$. 
This is the case for the ${\bf Z}_2$ orientifold model, where these considerations
apply. We found in last section that the tensor field $b$ couples
only to the gauge group of the 5 sector through $X_4^{(5)}$, whereas its Bianchi identity,
obtained by integrating by parts and dualizing the coupling to $\tilde b$, involves
only the 9 sector. Moreover $c_5=\tilde c_9>0$; no phase transitions
can occur in this case. 

Let us consider now the other orientifold models with $n_T>1$. 
The first task is to identify the combination of anti-self-dual and self-dual
tensor fields giving rise to the unconstrained untwisted $b$ field and to the remaining
$n_T-1$ tensors. Although these models do not have a clear geometric interpretation,
the kinetic term for the untwisted tensor field $b$, 
being independent of the details of the orientifold, can still be seen as a reduction
from $D=10$ to $D=6$ and has to be multiplied just by the volume $V_4$ of the 
internal orbifold.
In this case, we {\it define} the scalar $\phi$ such that $V_4= e^{-2 \langle \phi \rangle}$.
Let us now parametrize the scalar fields $v_r$ as $v_0=\cosh \phi$, $v_i=\Omega_i\sinh \phi$,
$i=1,...,n_T$, where $\Omega_i$ are coordinates on the unit $S^{(n_T-1)}$-sphere 
($\sum_i \Omega_i^2 = 1$).
With this choice of parametrization, there is a natural (and probably unique)
choice for the vacuum expectation values for the scalars $v_r$ and $x_r^M$, that 
reproduce the kinetic term for $H$: $\langle v \rangle_0 = \cosh \langle \phi \rangle
= \langle x \rangle_1^1$, $\langle v \rangle_1 = \sinh \langle \phi \rangle = 
\langle x \rangle_0^1$, $\langle x \rangle_i^I = \delta_i^I$ for $i,I=2,...,n_T$, all the 
other being equal to zero.
This strongly suggests that the orientifold string construction is appropriate
only at this point in the moduli space parametrized by the vacuum expectation
values of the scalars belonging to the extra tensor multiplets.
Notice that the vacuum $\langle\Omega_i\rangle=0$ breaks  
the global $SO(1,n_T)$ symmetry spontaneously to $SO(1,1)\times SO(n_T-1)$. 
The untwisted $b$ field is obtained precisely
like in the $n_T=1$ case whereas $b^i=K^{i(-)}$ are identified with the
additional $n_T-1$ anti-self-dual
two-forms. Correspondingly, at this point in moduli space the gauge coupling constants 
depend only on the
coefficients $c_z$ and $\tilde c_z$ associated to the Bianchi identity and couplings
of $b$ only. Notice that for all the ${\bf Z}_N$ models $c_9=\tilde c_5=0$ and
$c_5=\tilde c_9 > 0$. Again, all the gauge kinetic terms are positive definite at
this particular point in the moduli space.

All the coefficients $c_r^z$ do not depend on the vev's of the scalars
$v_r$ and can be fixed from the four-forms $X_4$ found in last section. 
This allows to analyze the gauge couplings for generic values of the moduli. 
We have checked that for each of the models 
considered here there exist a continuous family of points in moduli space 
where the gauge couplings vanish and tensionless strings occur. A detailed analysis of 
the loci where this phenomenon occurs and their interpretation would be extremely 
interesting.
As an illustrative example, consider the $\Z_3$ model. Modulo an irrelevant overall factor,
the constants $c^r_z$ for the two factors $z=a,b$ of the gauge group are found to be 
$c^0_a=-c^1_a = 1/2\sqrt{6}$, $c^0_b=-c^1_b = 1/4\sqrt{6}$, $c^m_a = 1/6\sqrt{2}$, 
$c^m_b = - 1/6\sqrt{2}$, $m=1,...,9$. It is then easy to check that for generic values
of $\langle \Omega_i \rangle$, there exists in general a value of $\langle \phi \rangle$,
corresponding to a particular $V_4$, for which $v_r c^r_z$ vanishes.
Choosing for instance 
$\langle \Omega_1 \rangle = 0$, $\langle \Omega_{i>1} \rangle = 1/3$, the
gauge couplings of $F_a$ and $F_b$ vanish for $\coth \langle \phi \rangle = \sqrt{3}$
and $\coth \langle \phi \rangle = 2\sqrt{3}$ respectively.

The linear couplings of the scalars to the field strengths can also be derived once 
the constants $c^r_z$ are known. Using the general results and notation of section four, 
these couplings can be rewritten in a concise way as
$$
S^{(k)}_{GC} =  \frac 1{32\pi^2} \sqrt{\frac {\pi}{N}} \sin \frac {\pi k}N \int \! d^6x  
\left\{ \chi^{(k)1} {\rm tr}\,[\gamma_k \, F_5^2] +
\Big(\frac 1{\sqrt{N_k}} \sum_{i_k=1}^{N_k} \chi^{(k)i_k}\Big) 
{\rm tr}\,[\gamma_k \, F_9^2] \right\} \;,
$$
where $\chi^{(k)}$, $k=1,...,N/2-1$ are the NSNS twisted scalars belonging to the 
additional tensor multiplets,
and $F_{5,9}$ are now conventional field-strengths rather than curvature two-forms.
As for the $9$ sector, the above couplings are in agreement with the recent results of 
\cite{abd}, modulo an overall numerical coefficient that we have not attempted to 
check\footnote{The scalar fields $m_k$ of \cite{abd} have to be identified as
$m_k = 1/\sqrt{N_k} \sum_{i_k=1}^{N_k} \chi^{(k)i_k}$.}.

\section{Discussion and conclusions}

In this paper, we have computed particular topological
amplitudes in the odd spin-structure enabling us to
understand in detail how anomalies cancel in a 
class of $N=1$ $D=6$ orientifold models with maximal
unbroken gauge group. By factorization, we have also found the
D-brane, O-plane and fixed-point couplings to the RR fields
arising in these models.

The mechanism of anomaly cancellation applies in its most general form,
with the exchange of different tensor fields \cite{sagGS} and,
whenever Abelian factors are present, scalar fields as well 
\cite{damo,americani}. The RR scalars involved in the inflow mechanism
belong always to neutral hypermultiplets and are responsible for a 
spontaneous symmetry breaking of all $U(1)$ factors but one 
in the ${\bf Z}_6$ model, in the maximally symmetric case.
More generically, the maximum number of $U(1)$ that can be
spontaneously broken never exceeds the number of neutral
hypermultiplets arising from the twisted closed string sector.

By comparing some terms in the low-energy effective actions of these
models with the most general ones allowed by supersymmetry, we deduce
that the vacuum expectation values of the scalars belonging to the extra 
tensor multiplets are fixed at particular values.
In these points in moduli space all the gauge kinetic terms are definite
positive. However, varying these moduli - flat directions in the 
low-energy action - we always find special loci where the gauge couplings
diverge and tensionless strings occur.
Analogously, we can also fix the gauge couplings of these scalars
for both the D9 and D5 gauge groups.

The models we discussed in this paper are particular examples
of six-dimensional theories with extra tensor multiplets. These arise for instance
in F and M-theory $D=6$ compactifications\footnote{See also \cite{ang} for other 
constructions of these vacua.}. In this last case, each extra tensor multiplet
is believed to come from an M5-brane in the bulk of the $11^{th}$ dimension \cite{witm5}.
Although the non-perturbative heterotic ${\rm Spin}(32)/{\bf Z}_2$ duals to the 
IIB orientifolds models are known \cite{afiuv}, their explicit realization as M-theory 
vacua is not a simple issue, especially in the maximally symmetric case discussed here. 
Nevertheless, some questions might still 
be qualitatively addressed. Assuming that each extra tensor multiplet comes from a single 
M5-brane, one might naively hope that these models are somehow related to M-theory  
compactified on an orbifold like $T^4/{\bf Z}_N\times S^1/{\bf Z}_2$.
However, from \cite{horwit} it is clear that such M-theory vacua corresponds to
non-perturbative vacua of strongly coupled $E_8 \times E_8$ heterotic theory on 
$T^4/{\bf Z}_N$, with total instanton number $25-n_T$. These vacua could in turn be related 
to $K3$ orientifolds. The $n_T-1$ M5-branes are located at distinct fixed-points of the 
orbifold. One is also tempted to consider somehow the vev 
$\langle \Omega_i \rangle $ as the moduli corresponding to move the $i^{th}$ brane on the 
$S^1/{\bf Z}_2$ segment. It seems that the orientifold description requires all these
M5-branes to be stuck at some point in the segment and at different fixed-points
on the $T^4/{\bf Z}_N$ orbifold.
On the other hand, in M-theory one has the freedom of moving the M5-branes along the segment
by turning on $\langle \Omega_i \rangle$. As already said, this leads to tensionless strings
at special points. It would be then very interesting to better understand
the connection with M-theory by a full and deeper analysis of this phenomenon.
In this direction, a more general analysis with different configurations
for the D5 branes and with Wilson lines turned on, would also be quite interesting.

Generalizing the idea of \cite{horwit}, in compactifying M-theory on orbifolds 
new twisted sectors are expected to arise at fixed-points. These will play a decisive role
in making the theory anomaly-free. It would be then extremely interesting
to see how new couplings to the twisted sectors, as well as the compactification of the 
$D=11$ Chern-Simons term $C_3\wedge X_8$ on orbifolds, are related to the anomalous 
couplings found here. This could yield important informations on M-theory at the microscopic
level.

\vskip 40pt


\acknowledgments

It is a pleasure to thank D. Ghoshal, A. Sagnotti and S. Theisen for 
enlightening discussions and very useful comments. We are also grateful to 
D. Bernardini, J. Pawelczyk and E. Scheidegger for interesting and continuous
exchange of ideas. C.A.S. also thanks the University of Amsterdam for hospitality.

This work has been supported by EEC under TMR contract ERBFMRX-CT96-0045 and 
by the Nederlandse Organisatie voor Wetenschappelijk Onderzoek (NWO).


\vspace{1cm}
\renewcommand{\theequation}{A.\arabic{equation}}
\setcounter{equation}{0}
\par \noindent{\large \bf A. Inflows}
\vspace{3mm}

In this appendix, we recall some formulae from \cite{cy} which are needed to
check that the anomalous couplings that we have proposed indeed reproduce the
inflows from which they have been derived.

\vskip 9pt
\noindent 
{\bf Sector ${\bf k=0}$}
\vskip 1pt 
\noindent
Two anomalous coupling of the form 
$$
S_A = \sqrt{2\pi} \int_A C_{(0)} \wedge A(F,R) \;,\;\;
S_B = \sqrt{2\pi} \int_B \tilde C_{(0)} \wedge B(F,R)
$$
with $C_{(0)}$, $\tilde C_{(0)}$ given by (\ref{C0}), 
produce an inflow on the AB intersection given by
$\delta_{\eta}\,(i\,S_{A+B})\,=\,2\pi i\,\int I_{AB}^{(1)}$
where 
\be
I_{AB} = \frac 12 \Big[A_4(F,R)\,B_4(-F,R) + A_4(-F,R)\,B_4(F,R)\Big] \;.
\ee
In this case all the couplings are even in $F$ due to the property 
${\rm ch}\,(-F) = {\rm ch}\,(F)$ of the Chern class in the Chan-Paton 
representation. The inflow then reduces to
\be
I_{AB} = A_4(F,R)\,B_4(F,R) = \Big[A(F,R) \wedge B(F,R)\Big]_8 \;.
\ee
Of course we neglect here and in the following the eight-form components 
$A_8(F,R)$ and $B_8(F,R)$, because their sum automatically vanishes once the tadpole
conditions are solved, and therefore there is no inflow associated with them.
For non-distinct objects, an additional $1/2$ is needed.

\vskip 9pt
\noindent 
{\bf Sector ${\bf k=N/2}$}
\vskip 1pt 
\noindent
Two anomalous coupling of the form 
$$
S_A = \sqrt{2\pi} \int_A C_{(N/2)}^i \wedge A(F,R) \;,\;\;
S_B = \sqrt{2\pi} \int_B C_{(N/2)}^i \wedge B(F,R)
$$
with $C^i_{(N/2)}$ given by (\ref{CN2}), 
produce an inflow on the AB intersection given by
\bea
I_{AB} \a=\a \frac 12 \Big[A_2(F,R)\,B_6(-F,R) + A_6(F,R)\,B_2(-F,R) \nn \\
\a\;\a \hspace{20pt} + A_2(-F,R)\,B_6(F,R) + A_6(-F,R)\,B_2(F,R)\Big] \;. 
\eea
Now all the couplings are odd in $F$ due to the property 
${\rm ch}\,(\gamma_{N/2}(-F)) = -{\rm ch}\,(\gamma_{N/2}F)$ in the Chan-Paton 
representation. The inflow then reduces to
\be
I_{AB} = -  A_2(F,R)\,B_6(F,R) -  A_6(F,R)\,B_2(F,R)
= - \Big[A(F,R) \wedge B(F,R)\Big]_8 \;.
\ee
For non-distinct objects, an additional $1/2$ is needed.

\vskip 9pt
\noindent 
{\bf Sectors ${\bf k \neq 0,N/2}$}
\vskip 1pt 
\noindent
The anomalous couplings
$$
S_A = \sqrt{2\pi} \int_A C_{(k)}^{i_k} \wedge A(F,R) \;,\;\;
S_B = \sqrt{2\pi} \int_B C_{(k)}^{i_k} \wedge B(F,R)
$$
with $C^{i_k}_{(k)}$ as in (\ref{Ck}), produce an inflow given by
\bea
I_{AB} \a=\a \frac 12 \Big[A_2(F,R)\,B_6(-F,R) + A_6(F,R)\,B_2(-F,R) \nn \\
\a\;\a \hspace{20pt} + A_2(-F,R)\,B_6(F,R) + A_6(-F,R)\,B_2(F,R) 
\raisebox{12pt}{} \nn \\
\a\;\a \hspace{20pt} - A_4(-F,R)\,B_4(F,R) - A_4(-F,R)\,B_4(F,R)\Big] 
\;. \raisebox{16pt}{}\ 
\eea
Furthermore, the 4-form couplings are even in $F$ whereas the 2 and 6-form couplings
are odd, due to the property ${\rm ch}\,(\gamma_{k}(-F))|_p = 
(-)^{p/2}{\rm ch}\,(\gamma_{k}F)|_p$ valid for the p-form component. 
The inflow then reduces to
\bea
I_{AB} \a=\a \, - \, A_2(F,R)\,B_6(F,R)\, -\, A_6(F,R)\,B_2(F,R) \, - \,
A_4(F,R)\,B_4(F,R) \nn \\
\a=\a \, - \Big[A(F,R) \wedge B(F,R)\Big]_8 \;. \raisebox{16pt}{} 
\eea
For non-distinct objects, an additional $1/2$ is needed as usual. 


\vspace{5mm}
\renewcommand{\theequation}{B.\arabic{equation}}
\setcounter{equation}{0}
\par \noindent{\large \bf B. Anomalous couplings}
\vspace{3mm}

We report here for completeness the Wess-Zumino couplings
computed in section five, but in a form that makes
clearer the different contribution due to D9-branes, D5-branes,
O9-planes, O5-planes and Fk fixed-points.

\vskip 9pt
\noindent 
{\bf $\Z_2$-model}
\vskip 1pt 
\noindent
There are 32 D9-branes, 32 D5-branes, 1 O9-plane and 16 O5-planes or F1 fixed-points
($\Z_2$-fixed). Applying the general results of section four, their anomalous 
couplings are found to be
\bea
\a\a S_{D5} = \sqrt{2 \pi} \int \!\left\{\frac 12 b \,{\rm ch}(F_5) 
- (\varphi + \tilde \varphi)^1 \,{\rm ch}(\gamma_1\,F_5)
\raisebox{20pt}{}\right\}\sqrt{\hat{A}(R)} \;, \nn \\
\a\a S_{D9} = \sqrt{2 \pi} \int \!\left\{\frac 12 \tilde b \,{\rm ch}(F_9) 
+ \frac 14 \sum_{j=1}^{16} (\varphi + \tilde \varphi)^j 
\,{\rm ch}(\gamma_2\,F_9)\right\}\sqrt{\hat{A}(R)} \;, \nn \\
\a\a S_{O5} = -\sqrt{2 \pi} \int  b \, \sqrt{\hat{L}(R)} 
\;, \raisebox{20pt}{} \nn \\
\a\a S_{O9} = -\sqrt{2 \pi} \int \tilde b \, \sqrt{\hat{L}(R)} 
\;. \raisebox{24pt}{} \label{AnoZ2}
\eea
It is understood that all the products are wedge products and that one has to pick 
up the six-form component of each term only.

\vskip 9pt
\noindent 
{\bf $\Z_3$-model}
\vskip 1pt 
\noindent
There are 32 D9-branes, 1 O9-plane and 9 F1 fixed-points ($\Z_3$-fixed),
with anomalous couplings given by
\bea
\a\a S_{D9} = \sqrt{2 \pi} \int \! 
\left\{\frac 16 \, \tilde b \, {\rm ch}(F_9) 
+ \frac 1{3\sqrt{2}}\sum_{m=1}^9 \Big[b^\prime \!- \tilde b^\prime
\!+\! \sqrt{2} \, (\phi^\prime \!+ \tilde \phi^\prime)\Big]^m 
\,{\rm ch}(\gamma_1\,F_9) \right\} \sqrt{\hat{A}(R)} \;, \nn \\
\a\a S_{O9} = -\sqrt{2 \pi} \int \! \frac 2{\sqrt{6}} \, \tilde b \, 
\sqrt{\hat{L}(R)} \;, \raisebox{20pt}{} \nn \\
\a\a S_{F_1} = -\sqrt{2 \pi} \int \! \frac 1{6\sqrt{2}} \sum_{m=1}^9 \,
(b^\prime - \tilde b^\prime)^m \, \sqrt{\hat{L}(R)} \;. \label{AnoZ3}
\eea

\vskip 9pt
\noindent 
{\bf $\Z_4$-model}
\vskip 1pt 
\noindent
There are 32 D9-branes, 32 D5-branes, 1 O9-plane, 16 O5-planes or F2-fixed-points
($\Z_2$-fixed), and 4 F1-fixed-points ($\Z_4$-fixed). Among the 16 scalars 
$\varphi^i_{(N/2)}$ of the $\Z_2$-twisted sector, only 10, $\varphi^j$, 
are physical. These are defined as $\varphi^j = \varphi^j_{(N/2)}$ for $j=1,...,4$, 
and $\varphi^j = \sqrt{2} \,(\varphi^j_{(N/2)} + \varphi^{j+6}_{(N/2)})$ for 
$j=5,...,10$. The anomalous couplings are then found to be
\bea
S_{D5} \a=\a \sqrt{2 \pi} \int \! \left\{\frac 1{2\sqrt{2}} \, b \,{\rm ch}(F_5) 
- \frac 12 \Big[b^\prime \!- \tilde b^\prime \!+\!
\sqrt{2} \,(\phi^\prime \!+ \tilde \phi^\prime)\Big]^1 \,{\rm ch}(\gamma_1\,F_5) 
\raisebox{20pt}{}\right. \nn \\ \a\;\a \hspace{40pt} \left.
- \frac 1{\sqrt{2}} (\varphi + \tilde \varphi)^1
{\rm ch}(\gamma_2\,F_5)\right\}\sqrt{\hat{A}(R)} \;, \nn \\
S_{D9} \a=\a \sqrt{2 \pi} \int \! \left\{\frac 1{2\sqrt{2}} \, \tilde b \,{\rm ch}(F_9) 
+ \frac 14 \sum_{m=1}^4 \Big[b^\prime \!- \tilde b^\prime \!+\! 
\sqrt{2} \,(\phi^\prime \!+ \tilde \phi^\prime)\Big]^m {\rm ch}(\gamma_1\,F_9) 
\raisebox{20pt}{} \right. \nn \\ \a\;\a \hspace{40pt} \left.
+ \frac 1{4\sqrt{2}} \Big[\sum_{j=1}^{4} (\varphi + \tilde \varphi)^j
\!+\! \sqrt{2}\sum_{j=5}^{10} (\varphi + \tilde \varphi)^j\Big]
\,{\rm ch}(\gamma_2\,F_9)\right\}\sqrt{\hat{A}(R)} \;, \nn \\
S_{O5} \a=\a -\sqrt{2 \pi} \int \! \frac 1{\sqrt{2}} \, b \, \sqrt{\hat{L}(R)} \;, \nn \\
S_{O9} \a=\a -\sqrt{2 \pi} \int \! \frac 1{\sqrt{2}} \, \tilde b \, \sqrt{\hat{L}(R)} 
\;, \nn \\
S_{F1} \a=\a 0 \raisebox{16pt}{} \;. \label{AnoZ4}
\eea

\vskip 9pt
\noindent 
{\bf $\Z_6$-model}
\vskip 1pt 
\noindent
There are 32 D9-branes, 32 D5-branes, 1 O9-plane, 16 O5-planes or F3 fixed-points
($\Z_2$-fixed), 1 F1 fixed-point ($\Z_6$-fixed), and 9 F2 fixed-points ($\Z_6$-fixed).
Among the 16 scalars $\varphi^i_{(N/2)}$ of the $\Z_2$-twisted sector, only 6, 
$\varphi^j$, are physical. These are defined as $\varphi^1 = \varphi^1_{(N/2)}$ 
and $\varphi^j = \sqrt{3} \,(\varphi^j_{(N/2)} + \varphi^{j+5}_{(N/2)} 
+ \varphi^{j+10}_{(N/2)})$ for $j=2,...,6$. 
Similarly, among the 9 copies of scalars and 2-forms $\phi^i_{(2)}$, $b^i_{(2)\mu\nu}$
of the $\Z_3$-twisted sector, only 5 copies, $\phi^{\prime\prime m}$, 
$b_{\mu\nu}^{\prime\prime m}$, are independent. These are defined
as $\phi^{\prime\prime 1} = \phi^1_{(N/2)}$, $b_{\mu\nu}^{\prime\prime 1} = 
b^1_{(2)\mu\nu}$, and 
$\phi^{\prime\prime m} = \sqrt{2}\,(\phi^m_{(2)} + \varphi^{m+4}_{(2)})$, 
$b_{\mu\nu}^{\prime\prime m} = \sqrt{2}\,(b^m_{(2)\mu\nu} + b^{m+4}_{(2)\mu\nu})$
for $m=2,...,5$. The anomalous couplings are then found to be
\bea
S_{D5} \a=\a \sqrt{2 \pi} \int \! \left\{\frac 1{2\sqrt{3}} \, b \,{\rm ch}(F_5) 
- \frac 1{2\sqrt{3}} \Big[b^\prime \!- \tilde b^\prime \!+\! 
\sqrt{2} \, (\phi^\prime \!+ \tilde \phi^\prime)\Big] \,{\rm ch}(\gamma_1\,F_5) 
\right. \nn \\ \a\;\a \hspace{38pt} \left.
- \frac 12 \Big[b^{\prime\prime} \!- \tilde b^{\prime\prime}
\!+\! \sqrt{2} \,(\phi^{\prime\prime} \!+ \tilde \phi^{\prime\prime})\Big]^1
\,{\rm ch}(\gamma_2\,F_5) \right. \nn \\ \a\;\a \hspace{38pt} \left.
- \frac 1{\sqrt{3}} (\varphi + \tilde \varphi)^1
{\rm ch}(\gamma_3\,F_5)\right\}\sqrt{\hat{A}(R)} \;, \nn \\
S_{D9} \a=\a \sqrt{2 \pi} \int \!\left\{\raisebox{20pt}{}
\frac 1{2\sqrt{3}} \, \tilde b \,{\rm ch}(F_9) 
+ \frac 1{2\sqrt{3}} \Big[b^\prime \!- \tilde b^\prime \!+\! 
\sqrt{2} \,(\phi^\prime \!+ \tilde \phi^\prime)\Big] \,{\rm ch}(\gamma_1\,F_9) 
\right. \nn \\ \a\;\a \hspace{38pt}
+ \frac 16 \Big[[b^{\prime\prime} \!- \tilde b^{\prime\prime} 
\!+\! \sqrt{2} \,(\phi^{\prime\prime} \!+ \tilde \phi^{\prime\prime})]^1 
\!+\! \sqrt{2} \! \sum_{m=2}^{5}[b^{\prime\prime} \!- \tilde b^{\prime\prime} 
\!+\! \sqrt{2} \,(\phi^{\prime\prime} \!+ \tilde \phi^{\prime\prime})]^m \Big] 
\,{\rm ch}(\gamma_2\,F_9) \raisebox{20pt}{} \hspace{-5pt} \nn \\ 
\a\;\a \hspace{38pt} \left. + \frac 1{4\sqrt{3}} 
\Big[(\varphi + \tilde \varphi)^1 \!+\! 
\sqrt{3} \sum_{j=2}^{6} (\varphi + \tilde \varphi)^j \Big]
\,{\rm ch}(\gamma_3\,F_9)\right\}\sqrt{\hat{A}(R)} \;, \nn \\
S_{O5} \a=\a -\sqrt{2 \pi} \int \! \frac 1{\sqrt{3}} \, b \, \sqrt{\hat{L}(R)} \;, \nn \\
S_{O9} \a=\a -\sqrt{2 \pi} \int \! \frac 1{\sqrt{3}} \, \tilde b \, \sqrt{\hat{L}(R)} 
\;,\nn \\
S_{F1} \a=\a -\sqrt{2 \pi} \int \frac 14 
(b^{\prime\prime} \!- \tilde b^{\prime\prime})^1 \sqrt{\hat{L}(R)} \;, \nn \\
S_{F2} \a=\a -\sqrt{2 \pi} \int \! - \frac 1{12} 
\Big[(b^{\prime\prime} \!- \tilde b^{\prime\prime})^1
\!+\! \sqrt{2} \! \sum_{m=2}^{5} (b^{\prime\prime} \!- \tilde b^{\prime\prime})^m \Big] 
\sqrt{\hat{L}(R)} \;. \label{AnoZ6}
\eea

\end{document}